\documentclass[amssymb,aps,prl,superscriptaddress,reprint,twocolumn]{revtex4-1}
\usepackage{graphicx}
\usepackage{mathtools}
\usepackage{siunitx}
\usepackage{xcolor}
\usepackage[normalem]{ulem}
\usepackage[inline]{enumitem}
\usepackage{hyperref}

\begin{document}

\title{Protein recruitment through indirect mechanochemical interactions}
\author{Andriy Goychuk}
\affiliation{Arnold Sommerfeld Center for Theoretical Physics and Center for NanoScience, Department of Physics, Ludwig-Maximilians-Universit\"at M\"unchen, Theresienstr. 37, D-80333 Munich, Germany}
\author{Erwin Frey}
\affiliation{Arnold Sommerfeld Center for Theoretical Physics and Center for NanoScience, Department of Physics, Ludwig-Maximilians-Universit\"at M\"unchen, Theresienstr. 37, D-80333 Munich, Germany}
\date{\today}

\begin{abstract}
Some of the key proteins essential for important cellular processes are capable of recruiting other proteins from the cytosol to phospholipid membranes. 
The physical basis for this cooperativity of binding is, surprisingly, still unclear. 
Here, we suggest a general feedback mechanism that explains cooperativity through mechanochemical coupling mediated by the mechanical properties of phospholipid membranes. 
Our theory predicts that protein recruitment, and therefore also protein pattern formation, involves membrane deformation, and is strongly affected by membrane composition.
\end{abstract}

\maketitle

Protein pattern formation is essential for the spatial organization of intracellular processes~\cite{Halatek2018}.
Examples of biological significance include Min oscillations that guide the positioning of the Z-ring to midcell in \textit{E.~coli}~\cite{Lutkenhaus2007}, the roles of cell polarization in determining the position of a new growth zone or bud site in \textit{S.~cerevisiae}~\cite{Johnson1999} and the anteroposterior axis of the embryo in \textit{C.~elegans}~\cite{Goldstein2007}, and spatiotemporal patterns formed by members of the Rho family of GTPases in eukaryotic cells~\cite{Lawson2018}.
Such self-organized patterns are the product of a dynamic interplay between diffusion (both in the cytosol and on the membrane) and biochemical reactions among proteins and between proteins and the membrane.  
A crucial motif in all of the biochemical reaction networks that drive these processes is a nonlinear feedback mechanism, which is generally termed \textit{recruitment}.
Here, membrane-bound proteins facilitate the binding of other soluble proteins from the cytosol to the membrane~\cite{Halatek2018}. 
For example, in \textit{E.~coli}, membrane-bound MinD is said to recruit both cytosolic MinD and MinE to the membrane.
What then is the physical basis for such cooperative binding between proteins and the membrane? 
One could adopt a purely chemical perspective and suggest an explanation based on classical concepts of binding cooperativity~\cite{Hill1913, Stefan2013}. 
However, an indiscriminately high chemical affinity between recruiting proteins would also promote protein aggregation in the cytosol as an unwanted side-effect. 
Then, to still facilitate specific recruitment to the membrane, a possible strategy is for individual proteins to change their conformation upon binding to the membrane so as to become chemically affine scaffolds for other proteins~\cite{Fischer2011,Encinar2013}.
In addition to these chemical interactions, binding of proteins to membranes inevitably invokes forces that can lead to membrane deformation.

\begin{figure}[!t]
	\includegraphics{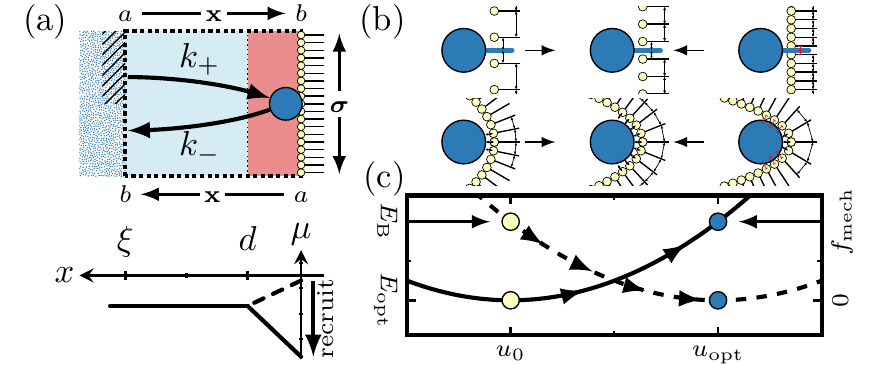}
	\caption[]{
	(a) We divide intracellular space into reaction compartments (top), each containing one protein on average (blue dot) and identify the distance from the membrane $x$ as the reaction coordinate.
	The proteins diffuse freely far away from the membrane ($x \,{>}\, d$, blue area) and sense a chemical potential $\mu$ close to the membrane ($x \,{<}\, d$, red area), which facilitates protein binding.
	Membrane-bound proteins modulate the chemical potential $\mu$ (arrow) and therefore induce a positive feedback in the attachment rate $k_+$.
	(b) Exaggerated membrane deformation illustrates protein interactions. Attachment occurs by (top) insertion of an anchor into the inner leaflet, or by (bottom) deposition through attractive surface interactions.
	(c) In both cases, the mechanical state change (arrows, $u \,{\in}\, \{\rho,H\}$) influences both the energy density $f_\text{mech}$ (solid line) stored in the deformation of the membrane and the binding energy of a protein $E_\text{B}$ (dashed line).	
	}
	\label{fig::illustration}
\end{figure}

Here we show how such mechanochemical coupling can lead to a mechanism for the cooperative recruitment of proteins to phospholipid membranes, and thereby provide an alternative strategy for cooperative membrane binding.
The basic idea is very simple:
Attractive forces between proteins and phospholipids facilitate protein attachment to the membrane.
As equal and opposite forces must act on the membrane as well, protein binding will induce mechanical deformation of the membrane.
Indeed, it is well known that membrane shape changes can be caused by curvature-inducing polymers and proteins~\cite{Ford2002, Tsafrir2003, Lee2005, Gov2006, Zimmerberg2006, Prinz2009, Stachowiak2012, McMahon2015, Jarsch2016, Gov2018} containing BAR-domains~\cite{Zimmerberg2004, Peter2004, Bhatia2009, Mim2012, Zhu2012, Prevost2015, Simunovic2015} and -- as recently shown~\cite{Litschel2018} -- also by the Min family of proteins.
Equilibrium theories of the coupling between proteins and membrane generally lead to membrane-mediated interactions between membrane-bound proteins, as reviewed in~\cite{Philips2009, Weikl2018, Idema2019}.
The physical origin of such interactions may be hydrophobic mismatch for integral proteins~\cite{Huang1986,Wiggins2005,Andersen2007,Milovanovic2015, GrauCampistany2015}, surface interactions that depend on curvature~\cite{Turner2004, Wiggins2005, Iglic2007, Shlomovitz2009, Perutkova2010, Zhu2012, Prevost2015, Mesarec2016, AgudoCanalejo2017}, or membrane shape fluctuations~\cite{Goulian1993, Golestanian1996}.
Furthermore, these interactions may also depend on the packing density~\cite{Schafer2011} and composition~\cite{Renner2012, Corradi2018} of the membrane.
Then, proteins that are bound to the membrane effectively attract or repel each other~\cite{Haselwandter2013, Schweitzer2015, vanderWel2016, Vahid2016}, and form different aggregates~\cite{Schmidt2008, Shlomovitz2009, Haselwandter2014, Mesarec2016, Vahid2017, Vahid2018, Idema2019}.
Here, however, we do not focus on such self-organization effects.
Instead, we ask a different and independent question, namely how membrane deformations affect the \emph{affinity and kinetic (un)binding rates} of proteins.
We propose a general protein recruitment mechanism caused by indirect interactions facilitated through mechanical deformations of the membrane.

As we are interested in quantifying the effect of membrane-mediated interactions on the \emph{kinetic rates of protein membrane binding and unbinding}, we need to analyze the \emph{dynamics of proteins} that are subject to both cytosolic diffusion (with diffusion constant $D$) and a chemical potential gradient $\mu ({\bf x})$ caused by the mechanochemical interaction of proteins with the membrane.
This is described by a Smoluchowski equation~\cite{Gardiner1986,Zwanzig2001} for the cytosolic protein density $c({\bf x}, t)$:
\begin{equation}
	\partial_t \, c ({\bf x}, t) 
	= 
	D \, {\boldsymbol \nabla}^2 c ({\bf x}, t)
	+ 
	\frac{D}{k_\text{B}T} \, {\boldsymbol \nabla} 
	\bigl( 
	c ({\bf x}, t) \, {\boldsymbol \nabla} \mu ({\bf x}) 
	\bigr) 
	\, .
\label{eq::smoluchowski}
\end{equation}
As proteins diffuse freely in the cytosol and interact with the membrane only within some narrow range $d$, a typical spatial profile of the chemical potential is initially flat in the cytosol ($\mu \,{=}\, 0$) and then monotonically approaches that of the proteins at the membrane, $\mu_\text{m} (m, {\bf u}) \,{=}\ \delta F [m ({\boldsymbol \sigma}), {\bf u} ({\boldsymbol \sigma})] / \delta m ({\boldsymbol \sigma})$, where $F$ denotes the free energy functional describing the mechanochemical interaction between proteins and membrane~\footnote{
In general, note that this implies that the chemical potential is a function of cytosolic position $\mathbf{x}$, and a functional of membrane protein density, $m(\boldsymbol\sigma)$. 
}.
In general, $F$ will depend on both the  membrane's protein density, $m ({\boldsymbol \sigma})$, and its mechanical state, ${\bf u} ({\boldsymbol \sigma})$, at position ${\boldsymbol \sigma}$ on the membrane surface;
see Fig.~\ref{fig::illustration} for an illustration.

The local free energy density describing the mechanochemical coupling between proteins and the membrane is determined by lipid-lipid and protein-lipid interactions. 
We assume that a fluid phospholipid membrane can, on a coarse-grained level, be considered as an elastically deformable thin sheet, with bulk modulus $\kappa_\text{s}$, vanishing shear modulus, and a bending modulus, $\kappa_\text{b}$, that is equal for both principal curvatures~\cite{Helfrich1973}.
For low levels of strain, we separate the mechanical degrees of freedom of the membrane into lateral stretching and out-of-plane bending~\cite{Seifert1997}, and write each mechanical contribution to the free energy as
\begin{equation}
	f_\text{mech} (u) 
	\,{=}\, 
	\tfrac12 
	\kappa \, (u \,{-}\, u_0)^2	\, .
\end{equation}
Here, $u \,{\in}\, \{ \rho, H \}$ is a placeholder variable for the mechanical state (conformation) of the membrane, $\kappa \,{\in}\, \{\kappa_\text{s}, \kappa_\text{b}\}$ denotes the respective membrane bulk and bending modulus, and $u_0$ denotes the equilibrium conformation (equilibrium density or intrinsic spontaneous curvature~\footnote{We further relate our approach to Helfrich's formulation of the bending energy cost~\cite{Helfrich1973} in the SM~\cite{supplement}.}).

As outlined above, there are several factors that determine the interaction between protein and membrane.
Conceptually, one may distinguish between two limiting cases [Fig.~\ref{fig::illustration}b]: 
\begin{enumerate*}[label={(\Alph*)}]
	\item Protein anchorage through a membrane-targeting domain that penetrates into the inner leaflet of the phospholipid bilayer and induces lateral membrane strain, \label{case::anchoring} \,\textit{or}\,
	\item protein attachment to the membrane by surface interactions and membrane bending. \label{case::deposition}
\end{enumerate*}
In both cases, the binding energy, $E_\text{B}\geq E_\text{opt}$, of a protein to the membrane will depend on the mechanical state (conformation) of the membrane, $u$.
In particular, the binding will be strongest, $E_\text{B} = E_\text{opt}$, for some optimal mechanical state, $u_\text{opt}$, where it attains an optimal value $E_\text{opt}<0$ [Fig.~\ref{fig::illustration}c].
This optimal conformation can be understood as a compromise between maximal attractive interactions between proteins and lipids, and minimal steric repulsion [Fig.~\ref{fig::illustration}b].
As the membrane becomes crowded with proteins, the binding energy will be reduced due to protein-protein interactions~\footnote{Note that the entropic effects of a large protein density can also reduce the protein binding energy, as discussed in the SM~\cite{supplement}. There, we show that the general result of nonlinear protein recruitment to the membrane remains valid.}. 
Given that the repulsive part of the Lennard-Jones potential scales as ${\propto}\, r^{-12}$ at small distances $r$, this may be accounted for by a factor $1 \,{+}\, \gamma \, m^6$ with $\gamma \,{<}\, 0$; note that the membrane protein density scales as $m \,{\propto}\, r^{-2}$.
Then, a Taylor expansion of the chemical free energy density to lowest order in the membrane conformation, $u$, yields 
\begin{equation}
	f_\text{chem} (u,m)
	=
	m \, 
	\bigl[
	E_\text{opt} \,(1+\gamma \, m^6)
	+
	\tfrac{1}{2} \, \epsilon \, 
	(u - u_\text{opt})^2
	\bigr] 
	\, ,
\label{eq::chemical_free_energy}
\end{equation}
where the parameter $\epsilon$ characterizes how strongly the membrane conformation affects protein binding.
As noted above, there is a broad range of cytosolic proteins that bind to lipid membranes in a curvature-dependent manner~\cite{Zimmerberg2004, Peter2004, Bhatia2009, Mim2012, Simunovic2015, McMahon2015, Zeno2018}; cf.\/ Fig.~\ref{fig::illustration}b, lower panel.
For example, protein-curvature coupling can arise from bending proteins to  the local membrane curvature~\cite{KraljIglic1999, Iglic2007, Perutkova2010, Shlomovitz2011, Bovic2015, Prevost2015, Mesarec2016}, or by bending the membrane to the shape of the proteins in order to maximize attractive interactions [Fig.~\ref{fig::illustration}b].
In the following, we specifically consider proteins that couple to the membrane curvature (sum of the two principal curvatures), $u\equiv H$, and discuss lipid-density-coupling proteins in the SM~\cite{supplement}.

As mechanical degrees of freedom relax much faster than protein densities, we adiabatically eliminate the mechanical degrees of freedom by assuming $\partial_u f \,{=}\, 0$, where $f=f_\text{mech}+f_\text{chem}$~\footnote{A further generalization yielding normal and tangential stresses involves variational surface calculus and is briefly outlined in the SM~\cite{supplement}. There, we show that the analysis presented here is valid in the limit of small deformations.}.
This yields a relation between the membrane conformation $u$ and the protein density $m$ on the membrane: $u(m) \,{=}\, u_0 \,{+}\, 
	\bigl( u_\text{opt} \,{-}\, u_0 \bigr) \, 
	{m}/(m_\times {+}\, m)$.
Here, the ratio between the mechanical modulus $\kappa$,  and the mechanochemical coupling parameter $\epsilon$, defines a characteristic membrane protein density: $m_\times \,{=}\, \kappa / \epsilon$.
For low membrane protein density, $m \,{<}\, m_\times$, the interaction between the lipids dominates, and the mechanical state of the membrane is given by the equilibrium value $u_0$; cf.\/ yellow symbols in Fig.~\ref{fig::illustration}c.
With increasing number of attached proteins, the membrane gradually deforms and adopts the mechanical state that is preferred by the proteins; cf.\/ blue symbols in Fig.~\ref{fig::illustration}c.
There is an interplay between a mechanical energy cost that is lowest at the relaxed state of the membrane, $u_0$, and a binding energy gain that is highest in the deformed state of the membrane which is optimal for protein binding, $u_\text{opt}$. 
The difference of mechanical free energy density and binding energy between the membrane conformations preferred by the proteins and the lipids read $\Delta f \,{\equiv}\, \Delta f_\text{mech} \,{=}\, \tfrac12 \kappa \,   (u_\text{opt} \,{-}\, u_0)^2$ and $\Delta E \,{\equiv}\, \Delta E_\text{B} \,{=}\, \tfrac12 \epsilon \, (u_\text{opt} \,{-}\, u_0)^2$, respectively.

\begin{figure}[bt]
\includegraphics{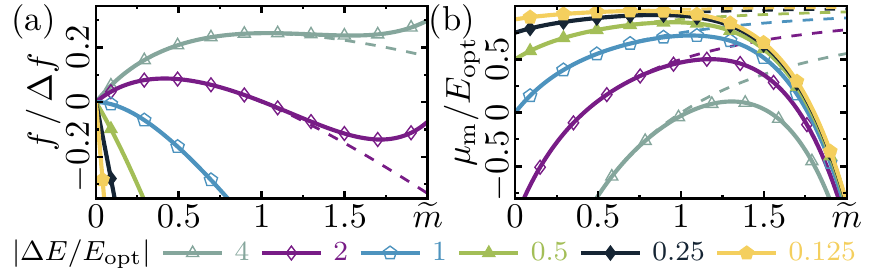}
\caption{
Free energy density (a), $f / \Delta f$,  and membrane chemical potential (b), $\mu_\text{m}/E_\text{opt}$, plotted as a function of the density of membrane-bound proteins, $m/m_\times$, for a series of different protein binding specificities, $|\Delta E/E_\text{\normalfont  opt}|$, indicated in the graph. 
Solid lines represent $\widetilde{\gamma} \,{=}\, -0.004$; dashed lines represent a system without crowding effects, $\widetilde{\gamma} \,{=}\, 0$.
}
\label{fig::free_energy}
\end{figure}

Upon eliminating the mechanical degrees of freedom using $u(m)$, the interplay between chemical and mechanical terms becomes obvious in the dependence of the free energy density on membrane protein density [Fig.~\ref{fig::free_energy}a],
\begin{equation}
	\frac{f}{\Delta f} 
	= 
	\frac{\widetilde{m}}{1 + \widetilde{m}} 
	+ 
	\widetilde{m} \,  
	\bigl(
	1 + \widetilde{\gamma} \, \widetilde{m}^6
	\bigr) \,
	\frac{E_\text{opt}}{\Delta E} 
	\, ,
\label{eq::free_energy_simple}
\end{equation}
where $\widetilde{m} \,{\coloneqq}\,  m/m_\times$ and $\widetilde{\gamma} \,{\coloneqq}\, \gamma / m_\times^6$.
The first term encodes free energy costs for membrane deformation through protein binding.
With increasing protein density, $m$, this contribution saturates, as the membrane deforms towards a binding-favorable conformation, implying that the corresponding mechanical free energy costs for binding of additional proteins diminish.
For intermediate membrane protein densities, the benefit from protein binding (second term in Eq.~\eqref{eq::free_energy_simple}) dominates.
Finally, for very high protein densities, protein binding becomes unfavorable due to crowding ($\widetilde{\gamma} \,{<}\, 0$).

The chemical potential at the membrane, $\mu_\text{m} \,{=}\, \partial_m f$, i.e.\/ the energy needed to bind one additional protein to the membrane, reads
\begin{equation}
	\frac{\mu_\text{m} (\widetilde{m})}
	     {E_\text{opt}} 
	= 
	1 + 
	7 \, \widetilde{\gamma} 
	  \, \widetilde{m}^{6}
	+
	\frac{\Delta E}{E_\text{opt}} \,
	\frac{1}{(1 + \widetilde{m})^2}   
	\, .
\label{eq::binding_energy_total}
\end{equation}
In the absence of crowding effects, the chemical potential approaches the optimal value $E_\text{opt} \,{<}\, 0$ for large protein densities on the membrane, $m \,{\gg}\, m_\times$, meaning that there is an energy gain upon binding [Fig.~\ref{fig::free_energy}b, dashed lines].
Crowding counteracts this gain, such that protein binding at high densities becomes unfavorable [Fig.~\ref{fig::free_energy}b, solid lines].
For low densities ($m \,{<}\, m_\times$), protein binding is also disfavored, as there is a free energy cost for mechanically deforming the membrane that is largest for low membrane protein densities $m$, cf.\/ the last term in Eq.~\eqref{eq::binding_energy_total}.
The amplitude of this reduction is given by $|\Delta E / E_\text{opt}|$, which we term the protein \textit{binding specificity}, as proteins with a higher specificity have a greater preference for mechanical states other than the relaxed state of the membrane [Fig.~\ref{fig::free_energy}b].
The less specific the binding of a protein, the smaller the changes in the chemical potential as a function of the protein density on the membrane.

What then are the implications of these thermodynamic considerations for the kinetics of protein binding and detachment?
To answer this question one has to solve a first-passage-time problem for a particle diffusing in a chemical potential as described by the Smoluchowski equation Eq.~\eqref{eq::smoluchowski}.
This is a well-studied problem, which dates back to Kramers' theory of reaction kinetics~\cite{Kramers1940}. 
For a one-dimensional reaction coordinate $x$, with a reflective boundary at $x \,{=}\, a$ and an absorbing boundary at $x \,{=}\, b$, the first-passage time is given by~\cite{Gardiner1986, Zwanzig2001}:
\begin{equation}
	\tau 
	= 
	\frac{1}{D} 
	\int_a^b \! dx \, 
	\text{e}^{+\mu(x)/k_\text{B}T} \, 
	\int_a^x \! dy \, 
	\text{e}^{-\mu(y)/k_\text{B}T} 
	\, ,
\label{eq::MFP}
\end{equation}
where $\mu (x)$ is the spatial profile of the chemical potential. 
In Kramers' classical escape problem, the reaction rate depends on the height of the barrier that the particle has to cross by diffusion to reach its target~\cite{Kramers1940}.
In our case, however, there is no such barrier. 
Instead, as discussed above, we expect the landscape to exhibit a monotonically increasing or decreasing profile, depending on whether the chemical potential at the membrane, $\mu_\text{m}$, is larger or smaller than the value in the bulk of the cytosol ($\mu_\text{cyt} \,{=}\, 0$); for an illustration see Fig.~\ref{fig::illustration}. 

To estimate the kinetic rates, we simplify the geometry of the cell as follows.
We divide the space near the membrane into small reaction compartments with respective sizes given by the average distance $\xi$ between proteins, such that each compartment contains a single protein on average.
Then, one may approximate a binding process as a one-dimensional diffusion process: an initially unbound protein diffusing in the cytosol enters one of these compartments at a distance $\xi$ from the membrane and after some time encounters the membrane located at $x \,{=}\, 0$.
To calculate the corresponding first-passage time, the membrane is considered as an absorbing boundary.
The cytosolic boundary of each compartment can effectively be approximated as a reflective boundary, since (on average) there is always one protein within each compartment, i.e.\/ a protein leaving the compartment at $x \,{=}\, \xi$ is replaced by one entering the compartment. 
Similarly, an unbinding process may be idealized as a stochastic process, where  an initially bound protein detaches at $x \,{=}\, 0$ (reflective boundary) and leaves the compartment at $x \,{=}\, \xi$ (absorbing boundary).

\begin{figure}[tb]
	\includegraphics{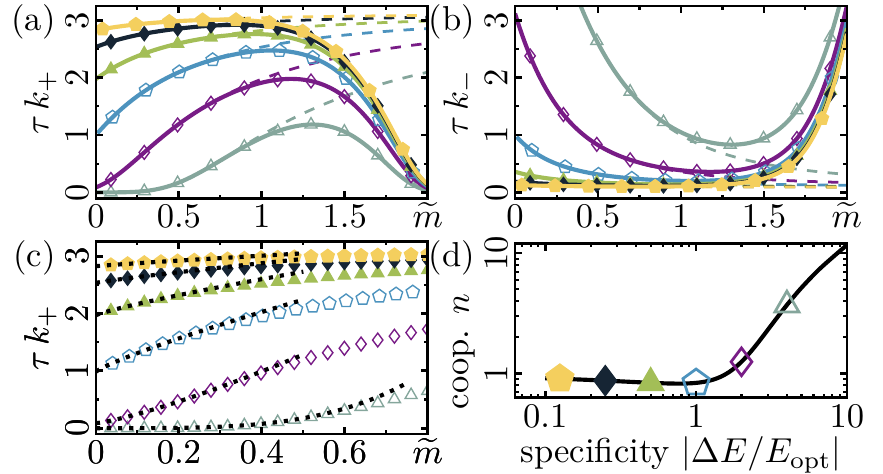}
	\caption{
	Kinetic rates for membrane attachment $k_+\, \tau$ (a) and detachment $k_- \, \tau$ (b) as a function of $\widetilde{m}$ for  $E_\text{opt} \,{=}\, {-}5 k_\text{\normalfont B} T$ and a set of protein specificities $|\Delta E / E_\text{opt}|$ as in Fig.~\ref{fig::free_energy}. Solid and dashed lines represent $\widetilde{\gamma} \,{=}\, {-}0.004$ and $\widetilde{\gamma} \,{=}\, 0$, respectively.
	(c) For low membrane protein concentrations $\widetilde{m}$, the attachment rate can be approximated by $a \,{+}\, b \, \widetilde{m}^n$; corresponding fits are indicated by the dotted lines.
	(d) The cooperativity coefficient $n$ increases with protein specificity $|\Delta E / E_\text{opt}|$.
	}
	\label{fig::binding_cooperativity}
\end{figure}

Given our limited knowledge of the profile of the chemical potential, we chose to approximate it by a piecewise linear function [Fig.~\ref{fig::illustration}a]. 
The protein diffuses freely ($\mu \,{=}\, 0$) at large distances from the membrane ($x \,{>}\, d$).
In the vicinity of the membrane ($x \,{<}\, d$),  we assume a linear profile $\mu \,{=}\, \mu_\text{m} (1 \,{-}\, x/d)$.
In the following, we discuss -- for simplicity -- the case where $\xi \,{=}\, d$.
The more general (and more realistic) case, where the protein also crosses a preceding flat potential of length $\xi \,{-}\, d \,{>}\, 0$, yields qualitatively similar results and is discussed in the SM~\cite{supplement}.
With these approximations, we can use  Eq.~\eqref{eq::MFP} to obtain an explicit analytic expression for the mean first-passage times $\tau_\pm$ of attachment and detachment~\cite{Bell2017}.
The corresponding kinetic rates, $k_\pm \,{=}\, \tau_\pm^{-1}$, expressed in units of the basic diffusion time $\tau \,{\coloneqq}\, 2 D/\xi^2$, are found to be
\begin{equation}
	k_\pm \tau 
	= 
	\tfrac{1}{2} 
	\Bigl(
	\tfrac{\mu_\text{m}}{k_\text{B} T}
	\Bigr)^2  
	\left( 
	\text{e}^{\pm \mu_\text{m}/k_\text{B} T} 
	\mp 
	\tfrac{\mu_\text{m}}{k_\text{B} T} 
	- 
	1 
	\right)^{-1} 
	\, .
\label{eq::rates}
\end{equation}
These rates exhibit a pronounced nonlinear dependence on the membrane protein density [Fig.~\ref{fig::binding_cooperativity}].
Hence, protein attachment and detachment are both cooperative processes, owing to the mechanochemical coupling mediated by membrane elasticity. 
By fitting the attachment rate, Eq.~\eqref{eq::rates}, at low densities with $k_+ \,{\approx}\, a \,{+}\, b \, \widetilde{m}^n$, we infer a relationship between the protein specificity $|\Delta E / E_\text{opt}|$ and the (Hill) cooperativity coefficient $n$ [Fig.~\ref{fig::binding_cooperativity}c,~d]; for an analysis in terms of Hill curves please refer to the SM~\cite{supplement}.  
Strong cooperativity ($n \,{>}\, 1$) occurs only for high protein specificities, $|\Delta E / E_\text{opt}| \,{>}\, 1$.
This implies that induction of a membrane conformation that favors protein binding requires the binding of a disproportionally large number of proteins to the membrane.
Therefore, in the deterministic limit, proteins would not attach to the membrane at all [Fig.~\ref{fig::free_energy}a, empty triangles and diamonds]. 
However, stochastic binding events, while unlikely at low protein densities, reduce the free energy cost of subsequent binding events and thereby increase their likelihood. 
This positive feedback leading to recruitment is a purely stochastic effect, and is related to nucleation during discontinuous phase transitions.

To assess whether the proposed indirect cooperativity mechanism could actually come into play at physiological protein concentrations, we estimated its various parameters from known literature values.
For proteins with a membrane sensing domain, typical values for the optimal curvature and binding energy are $H_\text{opt} \,{=}\, \SI{0.1}{\per\nano\meter}$~\cite{Peter2004, Wu2011} and  $E_\text{opt} \,{\approx}\, {-} 5 \, k_\text{B} T$~\cite{Lu2017}; we assume vanishing spontaneous curvature ($H_0 \,{=}\, 0$).
Across different studies, the bending modulus of a phospholipid bilayer was measured to be in the range of $\kappa \,{\approx}\, 10 \dots 50 \, k_\text{B} T$, suggesting a  typical value $\kappa \,{\approx}\, 30 k_\text{B} T$~\cite{Philips2009, Dimova2014, Nagle2015}.
Taking a value $|\Delta E / E_\text{opt}| \,{=}\, 2$ for protein specificity where nonlinear binding kinetics is significant (recruitment) [Fig.~\ref{fig::binding_cooperativity}d],
the corresponding range of concentrations, $m \,{<}\, m_\times \,{\approx}\, \SI{3e4}{\per\micro\meter\squared}$, easily encompasses any physiological value; the maximum packing density of proteins with size \SI{10}{\nano\meter} is $\SI{1e4}{\per\micro\meter\squared}$.

In summary, we have shown that mechanochemical coupling between proteins provides a possible mechanism for the nonlinear binding kinetics  (recruitment) of proteins to the membrane. 
The effect originates from the interplay between protein-lipid and lipid-lipid interactions, which induce mechanical deformations of the membrane and thereby alter the protein binding environment.
As protein-lipid interactions become dominant with increasing concentrations of membrane-bound proteins, the membrane's mechanical state becomes more favorable for binding.
This shows how cooperativity and the recruitment of proteins can naturally emerge without any reliance on direct chemical interactions and conformational changes.
The results should certainly be applicable to proteins that are known to bend membranes, e.g.\/ proteins containing BAR domains~\cite{Zimmerberg2004, Peter2004, Bhatia2009, Mim2012, Simunovic2015}.
As recent experiments have unexpectedly shown that Min protein oscillations can lead to oscillations in vesicle shape~\cite{Litschel2018}, we would argue that our theory should also apply to the broad class of NTPases that are essential for cellular protein pattern formation. 
Thus, strain sensing and generation might not only be a property of a few specialized proteins, but might actually be a prominent and perhaps general feature of membrane-binding proteins.
Further exploration of curvature sensing during macroscopic pattern formation might be highly rewarding~\cite{Peleg2011, Thalmeier2016, Wu2018}.
Our theory predicts that one can alter the recruitment exponent $n$ of membrane-binding proteins by tuning the protein specificity (possibly by changing the membrane composition or introducing permanently-bound membrane-bending proteins).
Such a change in cooperativity should have a much stronger effect on emerging protein patterns than the tuning of reaction rates, because it changes the nature of the nonlinear coupling.
We would expect profound changes in the protein dynamics that could be explored using appropriately modified reaction-diffusion models for various cellular systems~\cite{Huang2003, Halatek2012, Klunder2013, Denk2018, Goryachev2017, Halatek2018b}, as well as by experimentally tinkering with the composition of the membrane.
Finally, it would be highly interesting and rewarding to quantify the mechanochemical effect for specific membrane-binding proteins experimentally.
This would provide an interesting basis for theoretical models of pattern-forming protein systems and contribute towards revealing the universal role of membrane elasticity in cellular functions. \\

\goodbreak 

\begin{acknowledgments}
We thank Fridtjof Brauns, George Dadunashvili, Raphaela Ge\ss ele, Igor Goychuk, Isabella Graf, Laeschkir Hassan, Timon Idema, Anatoly B. Kolomeisky, Thomas Litschel, R\"udiger Thul and Manon Wigbers for stimulating discussions.
E.F. acknowledges financial support from the Deutsche Forschungsgemeinschaft (DFG) via the Collaborative Research Center (SFB) 1032 (project B2). 
A.G. is supported by a DFG fellowship through the Graduate School Quantitative Biosciences Munich (QBM). 
E.F. also acknowledges the hospitality of the Kavli Institute of Nanoscience at TU Delft, where part of this work was done.
\end{acknowledgments}


\begin{thebibliography}{87}%
\makeatletter
\providecommand \@ifxundefined [1]{%
 \@ifx{#1\undefined}
}%
\providecommand \@ifnum [1]{%
 \ifnum #1\expandafter \@firstoftwo
 \else \expandafter \@secondoftwo
 \fi
}%
\providecommand \@ifx [1]{%
 \ifx #1\expandafter \@firstoftwo
 \else \expandafter \@secondoftwo
 \fi
}%
\providecommand \natexlab [1]{#1}%
\providecommand \enquote  [1]{``#1''}%
\providecommand \bibnamefont  [1]{#1}%
\providecommand \bibfnamefont [1]{#1}%
\providecommand \citenamefont [1]{#1}%
\providecommand \href@noop [0]{\@secondoftwo}%
\providecommand \href [0]{\begingroup \@sanitize@url \@href}%
\providecommand \@href[1]{\@@startlink{#1}\@@href}%
\providecommand \@@href[1]{\endgroup#1\@@endlink}%
\providecommand \@sanitize@url [0]{\catcode `\\12\catcode `\$12\catcode
  `\&12\catcode `\#12\catcode `\^12\catcode `\_12\catcode `\%12\relax}%
\providecommand \@@startlink[1]{}%
\providecommand \@@endlink[0]{}%
\providecommand \url  [0]{\begingroup\@sanitize@url \@url }%
\providecommand \@url [1]{\endgroup\@href {#1}{\urlprefix }}%
\providecommand \urlprefix  [0]{URL }%
\providecommand \Eprint [0]{\href }%
\providecommand \doibase [0]{https://doi.org/}%
\providecommand \selectlanguage [0]{\@gobble}%
\providecommand \bibinfo  [0]{\@secondoftwo}%
\providecommand \bibfield  [0]{\@secondoftwo}%
\providecommand \translation [1]{[#1]}%
\providecommand \BibitemOpen [0]{}%
\providecommand \bibitemStop [0]{}%
\providecommand \bibitemNoStop [0]{.\EOS\space}%
\providecommand \EOS [0]{\spacefactor3000\relax}%
\providecommand \BibitemShut  [1]{\csname bibitem#1\endcsname}%
\let\auto@bib@innerbib\@empty
\bibitem [{\citenamefont {Halatek}\ \emph {et~al.}(2018)\citenamefont
  {Halatek}, \citenamefont {Brauns},\ and\ \citenamefont {Frey}}]{Halatek2018}%
  \BibitemOpen
  \bibfield  {author} {\bibinfo {author} {\bibfnamefont {J.}~\bibnamefont
  {Halatek}}, \bibinfo {author} {\bibfnamefont {F.}~\bibnamefont {Brauns}},\
  and\ \bibinfo {author} {\bibfnamefont {E.}~\bibnamefont {Frey}},\ }\href
  {https://doi.org/10.1098/rstb.2017.0107} {\bibfield  {journal} {\bibinfo
  {journal} {Philos. Trans. Royal Soc. B}\ }\textbf {\bibinfo {volume} {373}},\
  \bibinfo {pages} {20170107} (\bibinfo {year} {2018})}\BibitemShut {NoStop}%
\bibitem [{\citenamefont {Lutkenhaus}(2007)}]{Lutkenhaus2007}%
  \BibitemOpen
  \bibfield  {author} {\bibinfo {author} {\bibfnamefont {J.}~\bibnamefont
  {Lutkenhaus}},\ }\href
  {https://doi.org/10.1146/annurev.biochem.75.103004.142652} {\bibfield
  {journal} {\bibinfo  {journal} {Annu. Rev. Biochem.}\ }\textbf {\bibinfo
  {volume} {76}},\ \bibinfo {pages} {539} (\bibinfo {year} {2007})}\BibitemShut
  {NoStop}%
\bibitem [{\citenamefont {Johnson}(1999)}]{Johnson1999}%
  \BibitemOpen
  \bibfield  {author} {\bibinfo {author} {\bibfnamefont {D.~I.}\ \bibnamefont
  {Johnson}},\ }\href {https://mmbr.asm.org/content/63/1/54} {\bibfield
  {journal} {\bibinfo  {journal} {Microbiol. Mol. Biol. Rev.}\ }\textbf
  {\bibinfo {volume} {63}},\ \bibinfo {pages} {54} (\bibinfo {year}
  {1999})}\BibitemShut {NoStop}%
\bibitem [{\citenamefont {Goldstein}\ and\ \citenamefont
  {Macara}(2007)}]{Goldstein2007}%
  \BibitemOpen
  \bibfield  {author} {\bibinfo {author} {\bibfnamefont {B.}~\bibnamefont
  {Goldstein}}\ and\ \bibinfo {author} {\bibfnamefont {I.~G.}\ \bibnamefont
  {Macara}},\ }\href {https://doi.org/10.1016/j.devcel.2007.10.007} {\bibfield
  {journal} {\bibinfo  {journal} {Dev. Cell}\ }\textbf {\bibinfo {volume}
  {13}},\ \bibinfo {pages} {609} (\bibinfo {year} {2007})}\BibitemShut
  {NoStop}%
\bibitem [{\citenamefont {Lawson}\ and\ \citenamefont
  {Ridley}(2018)}]{Lawson2018}%
  \BibitemOpen
  \bibfield  {author} {\bibinfo {author} {\bibfnamefont {C.~D.}\ \bibnamefont
  {Lawson}}\ and\ \bibinfo {author} {\bibfnamefont {A.~J.}\ \bibnamefont
  {Ridley}},\ }\href {https://doi.org/10.1083/jcb.201612069} {\bibfield
  {journal} {\bibinfo  {journal} {J. Cell Biol.}\ }\textbf {\bibinfo {volume}
  {217}},\ \bibinfo {pages} {447} (\bibinfo {year} {2018})}\BibitemShut
  {NoStop}%
\bibitem [{\citenamefont {Hill}(1913)}]{Hill1913}%
  \BibitemOpen
  \bibfield  {author} {\bibinfo {author} {\bibfnamefont {A.~V.}\ \bibnamefont
  {Hill}},\ }\href {https://doi.org/10.1042/bj0070471} {\bibfield  {journal}
  {\bibinfo  {journal} {Biochem. J.}\ }\textbf {\bibinfo {volume} {7}},\
  \bibinfo {pages} {471} (\bibinfo {year} {1913})}\BibitemShut {NoStop}%
\bibitem [{\citenamefont {Stefan}\ and\ \citenamefont
  {Le~Nov\`{e}re}(2013)}]{Stefan2013}%
  \BibitemOpen
  \bibfield  {author} {\bibinfo {author} {\bibfnamefont {M.~I.}\ \bibnamefont
  {Stefan}}\ and\ \bibinfo {author} {\bibfnamefont {N.}~\bibnamefont
  {Le~Nov\`{e}re}},\ }\href {https://doi.org/10.1371/journal.pcbi.1003106}
  {\bibfield  {journal} {\bibinfo  {journal} {PLOS Comp. Biol.}\ }\textbf
  {\bibinfo {volume} {9}},\ \bibinfo {pages} {1} (\bibinfo {year}
  {2013})}\BibitemShut {NoStop}%
\bibitem [{\citenamefont {Fischer-Friedrich}\ and\ \citenamefont
  {Gov}(2011)}]{Fischer2011}%
  \BibitemOpen
  \bibfield  {author} {\bibinfo {author} {\bibfnamefont {E.}~\bibnamefont
  {Fischer-Friedrich}}\ and\ \bibinfo {author} {\bibfnamefont {N.}~\bibnamefont
  {Gov}},\ }\href {https://doi.org/10.1088/1478-3975/8/2/026007} {\bibfield
  {journal} {\bibinfo  {journal} {Phys. Biol.}\ }\textbf {\bibinfo {volume}
  {8}},\ \bibinfo {pages} {026007} (\bibinfo {year} {2011})}\BibitemShut
  {NoStop}%
\bibitem [{\citenamefont {Encinar}\ \emph {et~al.}(2013)\citenamefont
  {Encinar}, \citenamefont {Kralicek}, \citenamefont {Martos}, \citenamefont
  {Krupka}, \citenamefont {Cid}, \citenamefont {Alonso}, \citenamefont {Rico},
  \citenamefont {Jim{\'e}nez},\ and\ \citenamefont {V{\'e}lez}}]{Encinar2013}%
  \BibitemOpen
  \bibfield  {author} {\bibinfo {author} {\bibfnamefont {M.}~\bibnamefont
  {Encinar}}, \bibinfo {author} {\bibfnamefont {A.~V.}\ \bibnamefont
  {Kralicek}}, \bibinfo {author} {\bibfnamefont {A.}~\bibnamefont {Martos}},
  \bibinfo {author} {\bibfnamefont {M.}~\bibnamefont {Krupka}}, \bibinfo
  {author} {\bibfnamefont {S.}~\bibnamefont {Cid}}, \bibinfo {author}
  {\bibfnamefont {A.}~\bibnamefont {Alonso}}, \bibinfo {author} {\bibfnamefont
  {A.}~\bibnamefont {Rico}, \bibfnamefont {I.}}, \bibinfo {author}
  {\bibfnamefont {M.}~\bibnamefont {Jim{\'e}nez}},\ and\ \bibinfo {author}
  {\bibfnamefont {M.}~\bibnamefont {V{\'e}lez}},\ }\href
  {https://doi.org/10.1021/la401673z} {\bibfield  {journal} {\bibinfo
  {journal} {Langmuir}\ }\textbf {\bibinfo {volume} {29}},\ \bibinfo {pages}
  {9436} (\bibinfo {year} {2013})}\BibitemShut {NoStop}%
\bibitem [{\citenamefont {Ford}\ \emph {et~al.}(2002)\citenamefont {Ford},
  \citenamefont {Mills}, \citenamefont {Peter}, \citenamefont {Vallis},
  \citenamefont {Praefcke}, \citenamefont {Evans},\ and\ \citenamefont
  {McMahon}}]{Ford2002}%
  \BibitemOpen
  \bibfield  {author} {\bibinfo {author} {\bibfnamefont {M.~G.~J.}\
  \bibnamefont {Ford}}, \bibinfo {author} {\bibfnamefont {I.~G.}\ \bibnamefont
  {Mills}}, \bibinfo {author} {\bibfnamefont {B.~J.}\ \bibnamefont {Peter}},
  \bibinfo {author} {\bibfnamefont {Y.}~\bibnamefont {Vallis}}, \bibinfo
  {author} {\bibfnamefont {G.~J.~K.}\ \bibnamefont {Praefcke}}, \bibinfo
  {author} {\bibfnamefont {P.~R.}\ \bibnamefont {Evans}},\ and\ \bibinfo
  {author} {\bibfnamefont {H.~T.}\ \bibnamefont {McMahon}},\ }\href
  {https://doi.org/10.1038/nature01020} {\bibfield  {journal} {\bibinfo
  {journal} {Nature}\ }\textbf {\bibinfo {volume} {419}},\ \bibinfo {pages}
  {361} (\bibinfo {year} {2002})}\BibitemShut {NoStop}%
\bibitem [{\citenamefont {Tsafrir}\ \emph {et~al.}(2003)\citenamefont
  {Tsafrir}, \citenamefont {Caspi}, \citenamefont {Guedeau-Boudeville},
  \citenamefont {Arzi},\ and\ \citenamefont {Stavans}}]{Tsafrir2003}%
  \BibitemOpen
  \bibfield  {author} {\bibinfo {author} {\bibfnamefont {I.}~\bibnamefont
  {Tsafrir}}, \bibinfo {author} {\bibfnamefont {Y.}~\bibnamefont {Caspi}},
  \bibinfo {author} {\bibfnamefont {M.-A.}\ \bibnamefont {Guedeau-Boudeville}},
  \bibinfo {author} {\bibfnamefont {T.}~\bibnamefont {Arzi}},\ and\ \bibinfo
  {author} {\bibfnamefont {J.}~\bibnamefont {Stavans}},\ }\href
  {https://doi.org/10.1103/PhysRevLett.91.138102} {\bibfield  {journal}
  {\bibinfo  {journal} {Phys. Rev. Lett.}\ }\textbf {\bibinfo {volume} {91}},\
  \bibinfo {pages} {138102} (\bibinfo {year} {2003})}\BibitemShut {NoStop}%
\bibitem [{\citenamefont {Lee}\ \emph {et~al.}(2005)\citenamefont {Lee},
  \citenamefont {Orci}, \citenamefont {Hamamoto}, \citenamefont {Futai},
  \citenamefont {Ravazzola},\ and\ \citenamefont {Schekman}}]{Lee2005}%
  \BibitemOpen
  \bibfield  {author} {\bibinfo {author} {\bibfnamefont {M.~C.}\ \bibnamefont
  {Lee}}, \bibinfo {author} {\bibfnamefont {L.}~\bibnamefont {Orci}}, \bibinfo
  {author} {\bibfnamefont {S.}~\bibnamefont {Hamamoto}}, \bibinfo {author}
  {\bibfnamefont {E.}~\bibnamefont {Futai}}, \bibinfo {author} {\bibfnamefont
  {M.}~\bibnamefont {Ravazzola}},\ and\ \bibinfo {author} {\bibfnamefont
  {R.}~\bibnamefont {Schekman}},\ }\href
  {https://doi.org/10.1016/j.cell.2005.07.025} {\bibfield  {journal} {\bibinfo
  {journal} {Cell}\ }\textbf {\bibinfo {volume} {122}},\ \bibinfo {pages} {605
  } (\bibinfo {year} {2005})}\BibitemShut {NoStop}%
\bibitem [{\citenamefont {Gov}\ and\ \citenamefont
  {Gopinathan}(2006)}]{Gov2006}%
  \BibitemOpen
  \bibfield  {author} {\bibinfo {author} {\bibfnamefont {N.~S.}\ \bibnamefont
  {Gov}}\ and\ \bibinfo {author} {\bibfnamefont {A.}~\bibnamefont
  {Gopinathan}},\ }\href {https://doi.org/10.1529/biophysj.105.062224}
  {\bibfield  {journal} {\bibinfo  {journal} {Biophys. J.}\ }\textbf {\bibinfo
  {volume} {90}},\ \bibinfo {pages} {454 } (\bibinfo {year}
  {2006})}\BibitemShut {NoStop}%
\bibitem [{\citenamefont {Zimmerberg}\ and\ \citenamefont
  {Kozlov}(2006)}]{Zimmerberg2006}%
  \BibitemOpen
  \bibfield  {author} {\bibinfo {author} {\bibfnamefont {J.}~\bibnamefont
  {Zimmerberg}}\ and\ \bibinfo {author} {\bibfnamefont {M.~M.}\ \bibnamefont
  {Kozlov}},\ }\href@noop {} {\bibfield  {journal} {\bibinfo  {journal} {Nat.
  Rev. Mol. Cell Biol.}\ }\textbf {\bibinfo {volume} {7}},\ \bibinfo {pages} {9
  } (\bibinfo {year} {2006})}\BibitemShut {NoStop}%
\bibitem [{\citenamefont {Prinz}\ and\ \citenamefont
  {Hinshaw}(2009)}]{Prinz2009}%
  \BibitemOpen
  \bibfield  {author} {\bibinfo {author} {\bibfnamefont {W.~A.}\ \bibnamefont
  {Prinz}}\ and\ \bibinfo {author} {\bibfnamefont {J.~E.}\ \bibnamefont
  {Hinshaw}},\ }\href {https://doi.org/10.1080/10409230903183472} {\bibfield
  {journal} {\bibinfo  {journal} {Crit. Rev. Biochem. Mol. Biol.}\ }\textbf
  {\bibinfo {volume} {44}},\ \bibinfo {pages} {278} (\bibinfo {year}
  {2009})}\BibitemShut {NoStop}%
\bibitem [{\citenamefont {Stachowiak}\ \emph {et~al.}(2012)\citenamefont
  {Stachowiak}, \citenamefont {Schmid}, \citenamefont {Ryan}, \citenamefont
  {Ann}, \citenamefont {Sasaki}, \citenamefont {Sherman}, \citenamefont
  {Geissler}, \citenamefont {Fletcher},\ and\ \citenamefont
  {Hayden}}]{Stachowiak2012}%
  \BibitemOpen
  \bibfield  {author} {\bibinfo {author} {\bibfnamefont {J.~C.}\ \bibnamefont
  {Stachowiak}}, \bibinfo {author} {\bibfnamefont {E.~M.}\ \bibnamefont
  {Schmid}}, \bibinfo {author} {\bibfnamefont {C.~J.}\ \bibnamefont {Ryan}},
  \bibinfo {author} {\bibfnamefont {H.~S.}\ \bibnamefont {Ann}}, \bibinfo
  {author} {\bibfnamefont {D.~Y.}\ \bibnamefont {Sasaki}}, \bibinfo {author}
  {\bibfnamefont {M.~B.}\ \bibnamefont {Sherman}}, \bibinfo {author}
  {\bibfnamefont {P.~L.}\ \bibnamefont {Geissler}}, \bibinfo {author}
  {\bibfnamefont {D.~A.}\ \bibnamefont {Fletcher}},\ and\ \bibinfo {author}
  {\bibfnamefont {C.~C.}\ \bibnamefont {Hayden}},\ }\href@noop {} {\bibfield
  {journal} {\bibinfo  {journal} {Nat. Cell Biol.}\ }\textbf {\bibinfo {volume}
  {14}},\ \bibinfo {pages} {944–} (\bibinfo {year} {2012})}\BibitemShut
  {NoStop}%
\bibitem [{\citenamefont {McMahon}\ and\ \citenamefont
  {Boucrot}(2015)}]{McMahon2015}%
  \BibitemOpen
  \bibfield  {author} {\bibinfo {author} {\bibfnamefont {H.~T.}\ \bibnamefont
  {McMahon}}\ and\ \bibinfo {author} {\bibfnamefont {E.}~\bibnamefont
  {Boucrot}},\ }\href {https://doi.org/10.1242/jcs.114454} {\bibfield
  {journal} {\bibinfo  {journal} {J. Cell Sci.}\ }\textbf {\bibinfo {volume}
  {128}},\ \bibinfo {pages} {1065} (\bibinfo {year} {2015})}\BibitemShut
  {NoStop}%
\bibitem [{\citenamefont {Jarsch}\ \emph {et~al.}(2016)\citenamefont {Jarsch},
  \citenamefont {Daste},\ and\ \citenamefont {Gallop}}]{Jarsch2016}%
  \BibitemOpen
  \bibfield  {author} {\bibinfo {author} {\bibfnamefont {I.~K.}\ \bibnamefont
  {Jarsch}}, \bibinfo {author} {\bibfnamefont {F.}~\bibnamefont {Daste}},\ and\
  \bibinfo {author} {\bibfnamefont {J.~L.}\ \bibnamefont {Gallop}},\ }\href
  {https://doi.org/10.1083/jcb.201604003} {\bibfield  {journal} {\bibinfo
  {journal} {J. Cell Biol.}\ }\textbf {\bibinfo {volume} {214}},\ \bibinfo
  {pages} {375} (\bibinfo {year} {2016})}\BibitemShut {NoStop}%
\bibitem [{\citenamefont {Gov}(2018)}]{Gov2018}%
  \BibitemOpen
  \bibfield  {author} {\bibinfo {author} {\bibfnamefont {N.~S.}\ \bibnamefont
  {Gov}},\ }\href {https://doi.org/10.1098/rstb.2017.0115} {\bibfield
  {journal} {\bibinfo  {journal} {Philosophical Transactions of the Royal
  Society B: Biological Sciences}\ }\textbf {\bibinfo {volume} {373}},\
  \bibinfo {pages} {20170115} (\bibinfo {year} {2018})}\BibitemShut {NoStop}%
\bibitem [{\citenamefont {Zimmerberg}\ and\ \citenamefont
  {McLaughlin}(2004)}]{Zimmerberg2004}%
  \BibitemOpen
  \bibfield  {author} {\bibinfo {author} {\bibfnamefont {J.}~\bibnamefont
  {Zimmerberg}}\ and\ \bibinfo {author} {\bibfnamefont {S.}~\bibnamefont
  {McLaughlin}},\ }\href {https://doi.org/10.1016/j.cub.2004.02.060} {\bibfield
   {journal} {\bibinfo  {journal} {Curr. Biol.}\ }\textbf {\bibinfo {volume}
  {14}},\ \bibinfo {pages} {R250} (\bibinfo {year} {2004})}\BibitemShut
  {NoStop}%
\bibitem [{\citenamefont {Peter}\ \emph {et~al.}(2004)\citenamefont {Peter},
  \citenamefont {Kent}, \citenamefont {Mills}, \citenamefont {Vallis},
  \citenamefont {Butler}, \citenamefont {Evans},\ and\ \citenamefont
  {McMahon}}]{Peter2004}%
  \BibitemOpen
  \bibfield  {author} {\bibinfo {author} {\bibfnamefont {B.~J.}\ \bibnamefont
  {Peter}}, \bibinfo {author} {\bibfnamefont {H.~M.}\ \bibnamefont {Kent}},
  \bibinfo {author} {\bibfnamefont {I.~G.}\ \bibnamefont {Mills}}, \bibinfo
  {author} {\bibfnamefont {Y.}~\bibnamefont {Vallis}}, \bibinfo {author}
  {\bibfnamefont {P.~J.~G.}\ \bibnamefont {Butler}}, \bibinfo {author}
  {\bibfnamefont {P.~R.}\ \bibnamefont {Evans}},\ and\ \bibinfo {author}
  {\bibfnamefont {H.~T.}\ \bibnamefont {McMahon}},\ }\href
  {https://doi.org/10.1126/science.1092586} {\bibfield  {journal} {\bibinfo
  {journal} {Science}\ }\textbf {\bibinfo {volume} {303}},\ \bibinfo {pages}
  {495} (\bibinfo {year} {2004})}\BibitemShut {NoStop}%
\bibitem [{\citenamefont {Bhatia}\ \emph {et~al.}(2009)\citenamefont {Bhatia},
  \citenamefont {Madsen}, \citenamefont {Bolinger}, \citenamefont {Kunding},
  \citenamefont {Hedeg\r{a}rd}, \citenamefont {Gether},\ and\ \citenamefont
  {Stamou}}]{Bhatia2009}%
  \BibitemOpen
  \bibfield  {author} {\bibinfo {author} {\bibfnamefont {V.~K.}\ \bibnamefont
  {Bhatia}}, \bibinfo {author} {\bibfnamefont {K.~L.}\ \bibnamefont {Madsen}},
  \bibinfo {author} {\bibfnamefont {P.-Y.}\ \bibnamefont {Bolinger}}, \bibinfo
  {author} {\bibfnamefont {A.}~\bibnamefont {Kunding}}, \bibinfo {author}
  {\bibfnamefont {P.}~\bibnamefont {Hedeg\r{a}rd}}, \bibinfo {author}
  {\bibfnamefont {U.}~\bibnamefont {Gether}},\ and\ \bibinfo {author}
  {\bibfnamefont {D.}~\bibnamefont {Stamou}},\ }\href
  {https://doi.org/10.1038/emboj.2009.261} {\bibfield  {journal} {\bibinfo
  {journal} {EMBO J.}\ }\textbf {\bibinfo {volume} {28}},\ \bibinfo {pages}
  {3303} (\bibinfo {year} {2009})}\BibitemShut {NoStop}%
\bibitem [{\citenamefont {Mim}\ and\ \citenamefont {Unger}(2012)}]{Mim2012}%
  \BibitemOpen
  \bibfield  {author} {\bibinfo {author} {\bibfnamefont {C.}~\bibnamefont
  {Mim}}\ and\ \bibinfo {author} {\bibfnamefont {V.~M.}\ \bibnamefont
  {Unger}},\ }\href {https://doi.org/10.1016/j.tibs.2012.09.001} {\bibfield
  {journal} {\bibinfo  {journal} {Trends Biochem. Sci.}\ }\textbf {\bibinfo
  {volume} {37}},\ \bibinfo {pages} {526} (\bibinfo {year} {2012})}\BibitemShut
  {NoStop}%
\bibitem [{\citenamefont {Zhu}\ \emph {et~al.}(2012)\citenamefont {Zhu},
  \citenamefont {Das},\ and\ \citenamefont {Baumgart}}]{Zhu2012}%
  \BibitemOpen
  \bibfield  {author} {\bibinfo {author} {\bibfnamefont {C.}~\bibnamefont
  {Zhu}}, \bibinfo {author} {\bibfnamefont {S.~L.}\ \bibnamefont {Das}},\ and\
  \bibinfo {author} {\bibfnamefont {T.}~\bibnamefont {Baumgart}},\ }\href
  {https://doi.org/10.1016/j.bpj.2012.03.039} {\bibfield  {journal} {\bibinfo
  {journal} {Biophys. J.}\ }\textbf {\bibinfo {volume} {102}},\ \bibinfo
  {pages} {1837 } (\bibinfo {year} {2012})}\BibitemShut {NoStop}%
\bibitem [{\citenamefont {Pr{\'e}vost}\ \emph {et~al.}(2015)\citenamefont
  {Pr{\'e}vost}, \citenamefont {Zhao}, \citenamefont {Manzi}, \citenamefont
  {Lemichez}, \citenamefont {Lappalainen}, \citenamefont {Callan-Jones},\ and\
  \citenamefont {Bassereau}}]{Prevost2015}%
  \BibitemOpen
  \bibfield  {author} {\bibinfo {author} {\bibfnamefont {C.}~\bibnamefont
  {Pr{\'e}vost}}, \bibinfo {author} {\bibfnamefont {H.}~\bibnamefont {Zhao}},
  \bibinfo {author} {\bibfnamefont {J.}~\bibnamefont {Manzi}}, \bibinfo
  {author} {\bibfnamefont {E.}~\bibnamefont {Lemichez}}, \bibinfo {author}
  {\bibfnamefont {P.}~\bibnamefont {Lappalainen}}, \bibinfo {author}
  {\bibfnamefont {A.}~\bibnamefont {Callan-Jones}},\ and\ \bibinfo {author}
  {\bibfnamefont {P.}~\bibnamefont {Bassereau}},\ }\href
  {https://doi.org/10.1038/ncomms9529} {\bibfield  {journal} {\bibinfo
  {journal} {Nat. Commun.}\ }\textbf {\bibinfo {volume} {6}},\ \bibinfo {pages}
  {8529} (\bibinfo {year} {2015})}\BibitemShut {NoStop}%
\bibitem [{\citenamefont {Simunovic}\ \emph {et~al.}(2015)\citenamefont
  {Simunovic}, \citenamefont {Voth}, \citenamefont {Callan-Jones},\ and\
  \citenamefont {Bassereau}}]{Simunovic2015}%
  \BibitemOpen
  \bibfield  {author} {\bibinfo {author} {\bibfnamefont {M.}~\bibnamefont
  {Simunovic}}, \bibinfo {author} {\bibfnamefont {G.~A.}\ \bibnamefont {Voth}},
  \bibinfo {author} {\bibfnamefont {A.}~\bibnamefont {Callan-Jones}},\ and\
  \bibinfo {author} {\bibfnamefont {P.}~\bibnamefont {Bassereau}},\ }\href
  {https://doi.org/10.1016/j.tcb.2015.09.005} {\bibfield  {journal} {\bibinfo
  {journal} {Trends Cell Biol.}\ }\textbf {\bibinfo {volume} {25}},\ \bibinfo
  {pages} {780} (\bibinfo {year} {2015})}\BibitemShut {NoStop}%
\bibitem [{\citenamefont {Litschel}\ \emph {et~al.}(2018)\citenamefont
  {Litschel}, \citenamefont {Ramm}, \citenamefont {Maas}, \citenamefont
  {Heymann},\ and\ \citenamefont {Schwille}}]{Litschel2018}%
  \BibitemOpen
  \bibfield  {author} {\bibinfo {author} {\bibfnamefont {T.}~\bibnamefont
  {Litschel}}, \bibinfo {author} {\bibfnamefont {B.}~\bibnamefont {Ramm}},
  \bibinfo {author} {\bibfnamefont {R.}~\bibnamefont {Maas}}, \bibinfo {author}
  {\bibfnamefont {M.}~\bibnamefont {Heymann}},\ and\ \bibinfo {author}
  {\bibfnamefont {P.}~\bibnamefont {Schwille}},\ }\href
  {https://doi.org/10.1002/anie.201808750} {\bibfield  {journal} {\bibinfo
  {journal} {Angew. Chem. Int. Ed.}\ }\textbf {\bibinfo {volume} {57}},\
  \bibinfo {pages} {16286} (\bibinfo {year} {2018})}\BibitemShut {NoStop}%
\bibitem [{\citenamefont {Phillips}\ \emph {et~al.}(2009)\citenamefont
  {Phillips}, \citenamefont {Ursell}, \citenamefont {Wiggins},\ and\
  \citenamefont {Sens}}]{Philips2009}%
  \BibitemOpen
  \bibfield  {author} {\bibinfo {author} {\bibfnamefont {R.}~\bibnamefont
  {Phillips}}, \bibinfo {author} {\bibfnamefont {T.}~\bibnamefont {Ursell}},
  \bibinfo {author} {\bibfnamefont {P.}~\bibnamefont {Wiggins}},\ and\ \bibinfo
  {author} {\bibfnamefont {P.}~\bibnamefont {Sens}},\ }\href
  {https://doi.org/10.1038/nature08147} {\bibfield  {journal} {\bibinfo
  {journal} {Nature}\ }\textbf {\bibinfo {volume} {459}},\ \bibinfo {pages}
  {379} (\bibinfo {year} {2009})}\BibitemShut {NoStop}%
\bibitem [{\citenamefont {Weikl}(2018)}]{Weikl2018}%
  \BibitemOpen
  \bibfield  {author} {\bibinfo {author} {\bibfnamefont {T.~R.}\ \bibnamefont
  {Weikl}},\ }\href {https://doi.org/10.1146/annurev-physchem-052516-050637}
  {\bibfield  {journal} {\bibinfo  {journal} {Annu. Rev. Phys. Chem.}\ }\textbf
  {\bibinfo {volume} {69}},\ \bibinfo {pages} {521} (\bibinfo {year}
  {2018})}\BibitemShut {NoStop}%
\bibitem [{\citenamefont {Idema}\ and\ \citenamefont
  {Kraft}(2019)}]{Idema2019}%
  \BibitemOpen
  \bibfield  {author} {\bibinfo {author} {\bibfnamefont {T.}~\bibnamefont
  {Idema}}\ and\ \bibinfo {author} {\bibfnamefont {D.~J.}\ \bibnamefont
  {Kraft}},\ }\href {https://doi.org/10.1016/j.cocis.2019.01.006} {\bibfield
  {journal} {\bibinfo  {journal} {Curr. Opin. Colloid Interface Sci.}\ }\textbf
  {\bibinfo {volume} {40}},\ \bibinfo {pages} {58 } (\bibinfo {year}
  {2019})}\BibitemShut {NoStop}%
\bibitem [{\citenamefont {Huang}(1986)}]{Huang1986}%
  \BibitemOpen
  \bibfield  {author} {\bibinfo {author} {\bibfnamefont {H.}~\bibnamefont
  {Huang}},\ }\href {https://doi.org/10.1016/S0006-3495(86)83550-0} {\bibfield
  {journal} {\bibinfo  {journal} {Biophys. J.}\ }\textbf {\bibinfo {volume}
  {50}},\ \bibinfo {pages} {1061 } (\bibinfo {year} {1986})}\BibitemShut
  {NoStop}%
\bibitem [{\citenamefont {Wiggins}\ and\ \citenamefont
  {Phillips}(2005)}]{Wiggins2005}%
  \BibitemOpen
  \bibfield  {author} {\bibinfo {author} {\bibfnamefont {P.}~\bibnamefont
  {Wiggins}}\ and\ \bibinfo {author} {\bibfnamefont {R.}~\bibnamefont
  {Phillips}},\ }\href {https://doi.org/10.1529/biophysj.104.047431} {\bibfield
   {journal} {\bibinfo  {journal} {Biophys. J.}\ }\textbf {\bibinfo {volume}
  {88}},\ \bibinfo {pages} {880 } (\bibinfo {year} {2005})}\BibitemShut
  {NoStop}%
\bibitem [{\citenamefont {Andersen}\ and\ \citenamefont
  {Koeppe}(2007)}]{Andersen2007}%
  \BibitemOpen
  \bibfield  {author} {\bibinfo {author} {\bibfnamefont {O.~S.}\ \bibnamefont
  {Andersen}}\ and\ \bibinfo {author} {\bibfnamefont {R.~E.}\ \bibnamefont
  {Koeppe}},\ }\href {https://doi.org/10.1146/annurev.biophys.36.040306.132643}
  {\bibfield  {journal} {\bibinfo  {journal} {Annu. Rev. Biophys. Biomol.
  Struct.}\ }\textbf {\bibinfo {volume} {36}},\ \bibinfo {pages} {107}
  (\bibinfo {year} {2007})}\BibitemShut {NoStop}%
\bibitem [{\citenamefont {Milovanovic}\ \emph {et~al.}(2015)\citenamefont
  {Milovanovic}, \citenamefont {Honigmann}, \citenamefont {Koike},
  \citenamefont {G{\"o}ttfert}, \citenamefont {P{\"a}hler}, \citenamefont
  {Junius}, \citenamefont {M{\"u}llar}, \citenamefont {Diederichsen},
  \citenamefont {Janshoff}, \citenamefont {Grubm{\"u}ller}, \citenamefont
  {Risselada}, \citenamefont {Eggeling}, \citenamefont {Hell}, \citenamefont
  {van~den Bogaart},\ and\ \citenamefont {Jahn}}]{Milovanovic2015}%
  \BibitemOpen
  \bibfield  {author} {\bibinfo {author} {\bibfnamefont {D.}~\bibnamefont
  {Milovanovic}}, \bibinfo {author} {\bibfnamefont {A.}~\bibnamefont
  {Honigmann}}, \bibinfo {author} {\bibfnamefont {S.}~\bibnamefont {Koike}},
  \bibinfo {author} {\bibfnamefont {F.}~\bibnamefont {G{\"o}ttfert}}, \bibinfo
  {author} {\bibfnamefont {G.}~\bibnamefont {P{\"a}hler}}, \bibinfo {author}
  {\bibfnamefont {M.}~\bibnamefont {Junius}}, \bibinfo {author} {\bibfnamefont
  {S.}~\bibnamefont {M{\"u}llar}}, \bibinfo {author} {\bibfnamefont
  {U.}~\bibnamefont {Diederichsen}}, \bibinfo {author} {\bibfnamefont
  {A.}~\bibnamefont {Janshoff}}, \bibinfo {author} {\bibfnamefont
  {H.}~\bibnamefont {Grubm{\"u}ller}}, \bibinfo {author} {\bibfnamefont
  {H.~J.}\ \bibnamefont {Risselada}}, \bibinfo {author} {\bibfnamefont
  {C.}~\bibnamefont {Eggeling}}, \bibinfo {author} {\bibfnamefont {S.~W.}\
  \bibnamefont {Hell}}, \bibinfo {author} {\bibfnamefont {G.}~\bibnamefont
  {van~den Bogaart}},\ and\ \bibinfo {author} {\bibfnamefont {R.}~\bibnamefont
  {Jahn}},\ }\href {https://doi.org/10.1038/ncomms6984} {\bibfield  {journal}
  {\bibinfo  {journal} {Nature Comm.}\ }\textbf {\bibinfo {volume} {6}},\
  \bibinfo {pages} {5984} (\bibinfo {year} {2015})}\BibitemShut {NoStop}%
\bibitem [{\citenamefont {Grau-Campistany}\ \emph {et~al.}(2015)\citenamefont
  {Grau-Campistany}, \citenamefont {Strandberg}, \citenamefont {Wadhwani},
  \citenamefont {Reichert}, \citenamefont {B{\"u}rck}, \citenamefont
  {Rabanal},\ and\ \citenamefont {Ulrich}}]{GrauCampistany2015}%
  \BibitemOpen
  \bibfield  {author} {\bibinfo {author} {\bibfnamefont {A.}~\bibnamefont
  {Grau-Campistany}}, \bibinfo {author} {\bibfnamefont {E.}~\bibnamefont
  {Strandberg}}, \bibinfo {author} {\bibfnamefont {P.}~\bibnamefont
  {Wadhwani}}, \bibinfo {author} {\bibfnamefont {J.}~\bibnamefont {Reichert}},
  \bibinfo {author} {\bibfnamefont {J.}~\bibnamefont {B{\"u}rck}}, \bibinfo
  {author} {\bibfnamefont {F.}~\bibnamefont {Rabanal}},\ and\ \bibinfo {author}
  {\bibfnamefont {A.~S.}\ \bibnamefont {Ulrich}},\ }\href
  {https://doi.org/10.1038/srep09388} {\bibfield  {journal} {\bibinfo
  {journal} {Sci. Rep.}\ }\textbf {\bibinfo {volume} {5}},\ \bibinfo {pages}
  {9388} (\bibinfo {year} {2015})}\BibitemShut {NoStop}%
\bibitem [{\citenamefont {Turner}\ and\ \citenamefont
  {Sens}(2004)}]{Turner2004}%
  \BibitemOpen
  \bibfield  {author} {\bibinfo {author} {\bibfnamefont {M.~S.}\ \bibnamefont
  {Turner}}\ and\ \bibinfo {author} {\bibfnamefont {P.}~\bibnamefont {Sens}},\
  }\href {https://doi.org/10.1103/PhysRevLett.93.118103} {\bibfield  {journal}
  {\bibinfo  {journal} {Phys. Rev. Lett.}\ }\textbf {\bibinfo {volume} {93}},\
  \bibinfo {pages} {118103} (\bibinfo {year} {2004})}\BibitemShut {NoStop}%
\bibitem [{\citenamefont {Igli\v{c}}\ \emph {et~al.}(2007)\citenamefont
  {Igli\v{c}}, \citenamefont {Slivnik},\ and\ \citenamefont
  {Kralj-Igli\v{c}}}]{Iglic2007}%
  \BibitemOpen
  \bibfield  {author} {\bibinfo {author} {\bibfnamefont {A.}~\bibnamefont
  {Igli\v{c}}}, \bibinfo {author} {\bibfnamefont {T.}~\bibnamefont {Slivnik}},\
  and\ \bibinfo {author} {\bibfnamefont {V.}~\bibnamefont {Kralj-Igli\v{c}}},\
  }\href {https://doi.org/10.1016/j.jbiomech.2006.11.005} {\bibfield  {journal}
  {\bibinfo  {journal} {J. Biomech.}\ }\textbf {\bibinfo {volume} {40}},\
  \bibinfo {pages} {2492 } (\bibinfo {year} {2007})}\BibitemShut {NoStop}%
\bibitem [{\citenamefont {Shlomovitz}\ and\ \citenamefont
  {Gov}(2009)}]{Shlomovitz2009}%
  \BibitemOpen
  \bibfield  {author} {\bibinfo {author} {\bibfnamefont {R.}~\bibnamefont
  {Shlomovitz}}\ and\ \bibinfo {author} {\bibfnamefont {N.~S.}\ \bibnamefont
  {Gov}},\ }\href {https://doi.org/10.1088/1478-3975/6/4/046017} {\bibfield
  {journal} {\bibinfo  {journal} {Phys. Biol.}\ }\textbf {\bibinfo {volume}
  {6}},\ \bibinfo {pages} {046017} (\bibinfo {year} {2009})}\BibitemShut
  {NoStop}%
\bibitem [{\citenamefont {\v{S}\'{a}rka Perutkov\'{a}}\ \emph
  {et~al.}(2010)\citenamefont {\v{S}\'{a}rka Perutkov\'{a}}, \citenamefont
  {Kralj-Igli\v{c}}, \citenamefont {Frank},\ and\ \citenamefont
  {Igli\v{c}}}]{Perutkova2010}%
  \BibitemOpen
  \bibfield  {author} {\bibinfo {author} {\bibnamefont {\v{S}\'{a}rka
  Perutkov\'{a}}}, \bibinfo {author} {\bibfnamefont {V.}~\bibnamefont
  {Kralj-Igli\v{c}}}, \bibinfo {author} {\bibfnamefont {M.}~\bibnamefont
  {Frank}},\ and\ \bibinfo {author} {\bibfnamefont {A.}~\bibnamefont
  {Igli\v{c}}},\ }\href {https://doi.org/10} {\bibfield  {journal} {\bibinfo
  {journal} {J. Biomech.}\ }\textbf {\bibinfo {volume} {43}},\ \bibinfo {pages}
  {1612 } (\bibinfo {year} {2010})}\BibitemShut {NoStop}%
\bibitem [{\citenamefont {Mesarec}\ \emph {et~al.}(2016)\citenamefont
  {Mesarec}, \citenamefont {G\'{o}\'{z}d\'{z}}, \citenamefont {Igli\v{c}},
  \citenamefont {Kralj},\ and\ \citenamefont {Igli\v{c}}}]{Mesarec2016}%
  \BibitemOpen
  \bibfield  {author} {\bibinfo {author} {\bibfnamefont {L.}~\bibnamefont
  {Mesarec}}, \bibinfo {author} {\bibfnamefont {W.}~\bibnamefont
  {G\'{o}\'{z}d\'{z}}}, \bibinfo {author} {\bibfnamefont {V.~K.}\ \bibnamefont
  {Igli\v{c}}}, \bibinfo {author} {\bibfnamefont {S.}~\bibnamefont {Kralj}},\
  and\ \bibinfo {author} {\bibfnamefont {A.}~\bibnamefont {Igli\v{c}}},\ }\href
  {https://doi.org/10.1016/j.colsurfb.2016.01.010} {\bibfield  {journal}
  {\bibinfo  {journal} {Colloids Surf. B}\ }\textbf {\bibinfo {volume} {141}},\
  \bibinfo {pages} {132 } (\bibinfo {year} {2016})}\BibitemShut {NoStop}%
\bibitem [{\citenamefont {Agudo-Canalejo}\ and\ \citenamefont
  {Lipowsky}(2017)}]{AgudoCanalejo2017}%
  \BibitemOpen
  \bibfield  {author} {\bibinfo {author} {\bibfnamefont {J.}~\bibnamefont
  {Agudo-Canalejo}}\ and\ \bibinfo {author} {\bibfnamefont {R.}~\bibnamefont
  {Lipowsky}},\ }\href {https://doi.org/10.1039/C6SM02796B} {\bibfield
  {journal} {\bibinfo  {journal} {Soft Matter}\ }\textbf {\bibinfo {volume}
  {13}},\ \bibinfo {pages} {2155} (\bibinfo {year} {2017})}\BibitemShut
  {NoStop}%
\bibitem [{\citenamefont {Goulian}\ \emph {et~al.}(1993)\citenamefont
  {Goulian}, \citenamefont {Bruinsma},\ and\ \citenamefont
  {Pincus}}]{Goulian1993}%
  \BibitemOpen
  \bibfield  {author} {\bibinfo {author} {\bibfnamefont {M.}~\bibnamefont
  {Goulian}}, \bibinfo {author} {\bibfnamefont {R.}~\bibnamefont {Bruinsma}},\
  and\ \bibinfo {author} {\bibfnamefont {P.}~\bibnamefont {Pincus}},\ }\href
  {https://doi.org/10.1209/0295-5075/22/2/012} {\bibfield  {journal} {\bibinfo
  {journal} {Europhys. Lett.}\ }\textbf {\bibinfo {volume} {22}},\ \bibinfo
  {pages} {145} (\bibinfo {year} {1993})}\BibitemShut {NoStop}%
\bibitem [{\citenamefont {Golestanian}\ \emph {et~al.}(1996)\citenamefont
  {Golestanian}, \citenamefont {Goulian},\ and\ \citenamefont
  {Kardar}}]{Golestanian1996}%
  \BibitemOpen
  \bibfield  {author} {\bibinfo {author} {\bibfnamefont {R.}~\bibnamefont
  {Golestanian}}, \bibinfo {author} {\bibfnamefont {M.}~\bibnamefont
  {Goulian}},\ and\ \bibinfo {author} {\bibfnamefont {M.}~\bibnamefont
  {Kardar}},\ }\href {https://doi.org/10.1209/epl/i1996-00327-4} {\bibfield
  {journal} {\bibinfo  {journal} {Europhys. Lett.}\ }\textbf {\bibinfo {volume}
  {33}},\ \bibinfo {pages} {241} (\bibinfo {year} {1996})}\BibitemShut
  {NoStop}%
\bibitem [{\citenamefont {Sch{\"a}fer}\ \emph {et~al.}(2011)\citenamefont
  {Sch{\"a}fer}, \citenamefont {de~Jong}, \citenamefont {Holt}, \citenamefont
  {Rzepiela}, \citenamefont {de~Vries}, \citenamefont {Poolman}, \citenamefont
  {Killian},\ and\ \citenamefont {Marrink}}]{Schafer2011}%
  \BibitemOpen
  \bibfield  {author} {\bibinfo {author} {\bibfnamefont {L.~V.}\ \bibnamefont
  {Sch{\"a}fer}}, \bibinfo {author} {\bibfnamefont {D.~H.}\ \bibnamefont
  {de~Jong}}, \bibinfo {author} {\bibfnamefont {A.}~\bibnamefont {Holt}},
  \bibinfo {author} {\bibfnamefont {A.~J.}\ \bibnamefont {Rzepiela}}, \bibinfo
  {author} {\bibfnamefont {A.~H.}\ \bibnamefont {de~Vries}}, \bibinfo {author}
  {\bibfnamefont {B.}~\bibnamefont {Poolman}}, \bibinfo {author} {\bibfnamefont
  {J.~A.}\ \bibnamefont {Killian}},\ and\ \bibinfo {author} {\bibfnamefont
  {S.~J.}\ \bibnamefont {Marrink}},\ }\href
  {https://doi.org/10.1073/pnas.1009362108} {\bibfield  {journal} {\bibinfo
  {journal} {Proc. Natl. Acad. Sci. U.S.A.}\ }\textbf {\bibinfo {volume}
  {108}},\ \bibinfo {pages} {1343} (\bibinfo {year} {2011})}\BibitemShut
  {NoStop}%
\bibitem [{\citenamefont {Renner}\ and\ \citenamefont
  {Weibel}(2012)}]{Renner2012}%
  \BibitemOpen
  \bibfield  {author} {\bibinfo {author} {\bibfnamefont {L.~D.}\ \bibnamefont
  {Renner}}\ and\ \bibinfo {author} {\bibfnamefont {D.~B.}\ \bibnamefont
  {Weibel}},\ }\href {https://doi.org/10.1074/jbc.M112.407817} {\bibfield
  {journal} {\bibinfo  {journal} {J. Biol. Chem.}\ }\textbf {\bibinfo {volume}
  {287}},\ \bibinfo {pages} {38835} (\bibinfo {year} {2012})}\BibitemShut
  {NoStop}%
\bibitem [{\citenamefont {Corradi}\ \emph {et~al.}(2018)\citenamefont
  {Corradi}, \citenamefont {Mendez-Villuendas}, \citenamefont {Ing\'{o}lfsson},
  \citenamefont {Gu}, \citenamefont {Siuda}, \citenamefont {Melo},
  \citenamefont {Moussatova}, \citenamefont {DeGagn\'{e}}, \citenamefont
  {Sejdiu}, \citenamefont {Singh}, \citenamefont {Wassenaar}, \citenamefont
  {Delgado~Magnero}, \citenamefont {Marrink},\ and\ \citenamefont
  {Tieleman}}]{Corradi2018}%
  \BibitemOpen
  \bibfield  {author} {\bibinfo {author} {\bibfnamefont {V.}~\bibnamefont
  {Corradi}}, \bibinfo {author} {\bibfnamefont {E.}~\bibnamefont
  {Mendez-Villuendas}}, \bibinfo {author} {\bibfnamefont {H.~I.}\ \bibnamefont
  {Ing\'{o}lfsson}}, \bibinfo {author} {\bibfnamefont {R.-X.}\ \bibnamefont
  {Gu}}, \bibinfo {author} {\bibfnamefont {I.}~\bibnamefont {Siuda}}, \bibinfo
  {author} {\bibfnamefont {M.~N.}\ \bibnamefont {Melo}}, \bibinfo {author}
  {\bibfnamefont {A.}~\bibnamefont {Moussatova}}, \bibinfo {author}
  {\bibfnamefont {L.~J.}\ \bibnamefont {DeGagn\'{e}}}, \bibinfo {author}
  {\bibfnamefont {B.~I.}\ \bibnamefont {Sejdiu}}, \bibinfo {author}
  {\bibfnamefont {G.}~\bibnamefont {Singh}}, \bibinfo {author} {\bibfnamefont
  {T.~A.}\ \bibnamefont {Wassenaar}}, \bibinfo {author} {\bibfnamefont
  {K.}~\bibnamefont {Delgado~Magnero}}, \bibinfo {author} {\bibfnamefont
  {S.~J.}\ \bibnamefont {Marrink}},\ and\ \bibinfo {author} {\bibfnamefont
  {D.~P.}\ \bibnamefont {Tieleman}},\ }\href
  {https://doi.org/10.1021/acscentsci.8b00143} {\bibfield  {journal} {\bibinfo
  {journal} {ACS Cent. Sci.}\ }\textbf {\bibinfo {volume} {4}},\ \bibinfo
  {pages} {709} (\bibinfo {year} {2018})}\BibitemShut {NoStop}%
\bibitem [{\citenamefont {Haselwandter}\ and\ \citenamefont
  {Phillips}(2013)}]{Haselwandter2013}%
  \BibitemOpen
  \bibfield  {author} {\bibinfo {author} {\bibfnamefont {C.~A.}\ \bibnamefont
  {Haselwandter}}\ and\ \bibinfo {author} {\bibfnamefont {R.}~\bibnamefont
  {Phillips}},\ }\href {https://doi.org/10.1209/0295-5075/101/68002} {\bibfield
   {journal} {\bibinfo  {journal} {Europhys. Lett.}\ }\textbf {\bibinfo
  {volume} {101}},\ \bibinfo {pages} {68002} (\bibinfo {year}
  {2013})}\BibitemShut {NoStop}%
\bibitem [{\citenamefont {Schweitzer}\ and\ \citenamefont
  {Kozlov}(2015)}]{Schweitzer2015}%
  \BibitemOpen
  \bibfield  {author} {\bibinfo {author} {\bibfnamefont {Y.}~\bibnamefont
  {Schweitzer}}\ and\ \bibinfo {author} {\bibfnamefont {M.~M.}\ \bibnamefont
  {Kozlov}},\ }\href {https://doi.org/10.1371/journal.pcbi.1004054} {\bibfield
  {journal} {\bibinfo  {journal} {PLOS Comp. Biol.}\ }\textbf {\bibinfo
  {volume} {11}},\ \bibinfo {pages} {1} (\bibinfo {year} {2015})}\BibitemShut
  {NoStop}%
\bibitem [{\citenamefont {van~der Wel}\ \emph {et~al.}(2016)\citenamefont
  {van~der Wel}, \citenamefont {Vahid}, \citenamefont {\v{S}ari\'{c}},
  \citenamefont {Idema}, \citenamefont {Heinrich},\ and\ \citenamefont
  {Kraft}}]{vanderWel2016}%
  \BibitemOpen
  \bibfield  {author} {\bibinfo {author} {\bibfnamefont {C.}~\bibnamefont
  {van~der Wel}}, \bibinfo {author} {\bibfnamefont {A.}~\bibnamefont {Vahid}},
  \bibinfo {author} {\bibfnamefont {A.}~\bibnamefont {\v{S}ari\'{c}}}, \bibinfo
  {author} {\bibfnamefont {T.}~\bibnamefont {Idema}}, \bibinfo {author}
  {\bibfnamefont {D.}~\bibnamefont {Heinrich}},\ and\ \bibinfo {author}
  {\bibfnamefont {D.~J.}\ \bibnamefont {Kraft}},\ }\href
  {https://doi.org/10.1038/srep32825} {\bibfield  {journal} {\bibinfo
  {journal} {Sci. Rep.}\ }\textbf {\bibinfo {volume} {6}},\ \bibinfo {pages}
  {32825} (\bibinfo {year} {2016})}\BibitemShut {NoStop}%
\bibitem [{\citenamefont {Vahid}\ and\ \citenamefont
  {Idema}(2016)}]{Vahid2016}%
  \BibitemOpen
  \bibfield  {author} {\bibinfo {author} {\bibfnamefont {A.}~\bibnamefont
  {Vahid}}\ and\ \bibinfo {author} {\bibfnamefont {T.}~\bibnamefont {Idema}},\
  }\href {https://doi.org/10.1103/PhysRevLett.117.138102} {\bibfield  {journal}
  {\bibinfo  {journal} {Phys. Rev. Lett.}\ }\textbf {\bibinfo {volume} {117}},\
  \bibinfo {pages} {138102} (\bibinfo {year} {2016})}\BibitemShut {NoStop}%
\bibitem [{\citenamefont {Schmidt}\ \emph {et~al.}(2008)\citenamefont
  {Schmidt}, \citenamefont {Guigas},\ and\ \citenamefont
  {Weiss}}]{Schmidt2008}%
  \BibitemOpen
  \bibfield  {author} {\bibinfo {author} {\bibfnamefont {U.}~\bibnamefont
  {Schmidt}}, \bibinfo {author} {\bibfnamefont {G.}~\bibnamefont {Guigas}},\
  and\ \bibinfo {author} {\bibfnamefont {M.}~\bibnamefont {Weiss}},\ }\href
  {https://doi.org/10.1103/PhysRevLett.101.128104} {\bibfield  {journal}
  {\bibinfo  {journal} {Phys. Rev. Lett.}\ }\textbf {\bibinfo {volume} {101}},\
  \bibinfo {pages} {128104} (\bibinfo {year} {2008})}\BibitemShut {NoStop}%
\bibitem [{\citenamefont {Haselwandter}\ and\ \citenamefont
  {Wingreen}(2014)}]{Haselwandter2014}%
  \BibitemOpen
  \bibfield  {author} {\bibinfo {author} {\bibfnamefont {C.~A.}\ \bibnamefont
  {Haselwandter}}\ and\ \bibinfo {author} {\bibfnamefont {N.~S.}\ \bibnamefont
  {Wingreen}},\ }\href {https://doi.org/10.1371/journal.pcbi.1003932}
  {\bibfield  {journal} {\bibinfo  {journal} {PLOS Comp. Biol.}\ }\textbf
  {\bibinfo {volume} {10}},\ \bibinfo {pages} {1} (\bibinfo {year}
  {2014})}\BibitemShut {NoStop}%
\bibitem [{\citenamefont {Vahid}\ \emph {et~al.}(2017)\citenamefont {Vahid},
  \citenamefont {\v{S}ari\'{c}},\ and\ \citenamefont {Idema}}]{Vahid2017}%
  \BibitemOpen
  \bibfield  {author} {\bibinfo {author} {\bibfnamefont {A.}~\bibnamefont
  {Vahid}}, \bibinfo {author} {\bibfnamefont {A.}~\bibnamefont
  {\v{S}ari\'{c}}},\ and\ \bibinfo {author} {\bibfnamefont {T.}~\bibnamefont
  {Idema}},\ }\href {https://doi.org/10.1039/C7SM00433H} {\bibfield  {journal}
  {\bibinfo  {journal} {Soft Matter}\ }\textbf {\bibinfo {volume} {13}},\
  \bibinfo {pages} {4924} (\bibinfo {year} {2017})}\BibitemShut {NoStop}%
\bibitem [{\citenamefont {Vahid}\ and\ \citenamefont
  {Idema}(2018)}]{Vahid2018}%
  \BibitemOpen
  \bibfield  {author} {\bibinfo {author} {\bibfnamefont {A.}~\bibnamefont
  {Vahid}}\ and\ \bibinfo {author} {\bibfnamefont {T.}~\bibnamefont {Idema}},\
  }\bibfield  {journal} {\bibinfo  {journal} {bioRxiv}\ }\href
  {https://doi.org/10.1101/336545} {10.1101/336545} (\bibinfo {year}
  {2018})\BibitemShut {NoStop}%
\bibitem [{\citenamefont {Gardiner}(2009)}]{Gardiner1986}%
  \BibitemOpen
  \bibfield  {author} {\bibinfo {author} {\bibfnamefont {C.}~\bibnamefont
  {Gardiner}},\ }\href@noop {} {\emph {\bibinfo {title} {Stochastic Methods}}}\
  (\bibinfo  {publisher} {Springer Berlin Heidelberg},\ \bibinfo {year}
  {2009})\BibitemShut {NoStop}%
\bibitem [{\citenamefont {Zwanzig}(2001)}]{Zwanzig2001}%
  \BibitemOpen
  \bibfield  {author} {\bibinfo {author} {\bibfnamefont {R.}~\bibnamefont
  {Zwanzig}},\ }\href@noop {} {\emph {\bibinfo {title} {Nonequilibrium
  Statistical Mechanics}}}\ (\bibinfo  {publisher} {Oxford University Press},\
  \bibinfo {year} {2001})\BibitemShut {NoStop}%
\bibitem [{Note1()}]{Note1}%
  \BibitemOpen
  \bibinfo {note} {In general, note that this implies that the chemical
  potential is a function of cytosolic position $\protect \mathbf {x}$, and a
  functional of membrane protein density, $m(\protect \boldsymbol \sigma
  )$.}\BibitemShut {Stop}%
\bibitem [{\citenamefont {Helfrich}(1973)}]{Helfrich1973}%
  \BibitemOpen
  \bibfield  {author} {\bibinfo {author} {\bibfnamefont {W.}~\bibnamefont
  {Helfrich}},\ }\href@noop {} {\bibfield  {journal} {\bibinfo  {journal} {Z.
  Naturforsch. C Bio. Sci.}\ }\textbf {\bibinfo {volume} {28}},\ \bibinfo
  {pages} {693} (\bibinfo {year} {1973})}\BibitemShut {NoStop}%
\bibitem [{\citenamefont {Seifert}(1997)}]{Seifert1997}%
  \BibitemOpen
  \bibfield  {author} {\bibinfo {author} {\bibfnamefont {U.}~\bibnamefont
  {Seifert}},\ }\href {https://doi.org/10.1080/00018739700101488} {\bibfield
  {journal} {\bibinfo  {journal} {Adv. Phys.}\ }\textbf {\bibinfo {volume}
  {46}},\ \bibinfo {pages} {13} (\bibinfo {year} {1997})}\BibitemShut {NoStop}%
\bibitem [{Note2()}]{Note2}%
  \BibitemOpen
  \bibinfo {note} {We further relate our approach to Helfrich's formulation of
  the bending energy cost~\cite {Helfrich1973} in the SM~\cite
  {supplement}.}\BibitemShut {Stop}%
\bibitem [{Note3()}]{Note3}%
  \BibitemOpen
  \bibinfo {note} {Note that the entropic effects of a large protein density
  can also reduce the protein binding energy, as discussed in the SM~\cite
  {supplement}. There, we show that the general result of nonlinear protein
  recruitment to the membrane remains valid.}\BibitemShut {Stop}%
\bibitem [{\citenamefont {Zeno}\ \emph {et~al.}(2018)\citenamefont {Zeno},
  \citenamefont {Baul}, \citenamefont {Snead}, \citenamefont {DeGroot},
  \citenamefont {Wang}, \citenamefont {Lafer}, \citenamefont {Thirumalai},\
  and\ \citenamefont {Stachowiak}}]{Zeno2018}%
  \BibitemOpen
  \bibfield  {author} {\bibinfo {author} {\bibfnamefont {W.~F.}\ \bibnamefont
  {Zeno}}, \bibinfo {author} {\bibfnamefont {U.}~\bibnamefont {Baul}}, \bibinfo
  {author} {\bibfnamefont {W.~T.}\ \bibnamefont {Snead}}, \bibinfo {author}
  {\bibfnamefont {A.~C.~M.}\ \bibnamefont {DeGroot}}, \bibinfo {author}
  {\bibfnamefont {L.}~\bibnamefont {Wang}}, \bibinfo {author} {\bibfnamefont
  {E.~M.}\ \bibnamefont {Lafer}}, \bibinfo {author} {\bibfnamefont
  {D.}~\bibnamefont {Thirumalai}},\ and\ \bibinfo {author} {\bibfnamefont
  {J.~C.}\ \bibnamefont {Stachowiak}},\ }\href
  {https://doi.org/10.1038/s41467-018-06532-3} {\bibfield  {journal} {\bibinfo
  {journal} {Nature Comm.}\ }\textbf {\bibinfo {volume} {9}},\ \bibinfo {pages}
  {4152} (\bibinfo {year} {2018})}\BibitemShut {NoStop}%
\bibitem [{\citenamefont {Kralj-Igli{\v{c}}}\ \emph {et~al.}(1999)\citenamefont
  {Kralj-Igli{\v{c}}}, \citenamefont {Heinrich}, \citenamefont {Svetina},\ and\
  \citenamefont {{\v{Z}}ek{\v{s}}}}]{KraljIglic1999}%
  \BibitemOpen
  \bibfield  {author} {\bibinfo {author} {\bibfnamefont {V.}~\bibnamefont
  {Kralj-Igli{\v{c}}}}, \bibinfo {author} {\bibfnamefont {V.}~\bibnamefont
  {Heinrich}}, \bibinfo {author} {\bibfnamefont {S.}~\bibnamefont {Svetina}},\
  and\ \bibinfo {author} {\bibfnamefont {B.}~\bibnamefont {{\v{Z}}ek{\v{s}}}},\
  }\href {https://doi.org/10.1007/s100510050822} {\bibfield  {journal}
  {\bibinfo  {journal} {Eur. Phys. J. B}\ }\textbf {\bibinfo {volume} {10}},\
  \bibinfo {pages} {5} (\bibinfo {year} {1999})}\BibitemShut {NoStop}%
\bibitem [{\citenamefont {Shlomovitz}\ \emph {et~al.}(2011)\citenamefont
  {Shlomovitz}, \citenamefont {Gov},\ and\ \citenamefont
  {Roux}}]{Shlomovitz2011}%
  \BibitemOpen
  \bibfield  {author} {\bibinfo {author} {\bibfnamefont {R.}~\bibnamefont
  {Shlomovitz}}, \bibinfo {author} {\bibfnamefont {N.~S.}\ \bibnamefont
  {Gov}},\ and\ \bibinfo {author} {\bibfnamefont {A.}~\bibnamefont {Roux}},\
  }\href {https://doi.org/10.1088/1367-2630/13/6/065008} {\bibfield  {journal}
  {\bibinfo  {journal} {New J. Phys.}\ }\textbf {\bibinfo {volume} {13}},\
  \bibinfo {pages} {065008} (\bibinfo {year} {2011})}\BibitemShut {NoStop}%
\bibitem [{\citenamefont {Bov\v{z}i\v{c}}\ \emph {et~al.}(2015)\citenamefont
  {Bov\v{z}i\v{c}}, \citenamefont {Das},\ and\ \citenamefont
  {Svetina}}]{Bovic2015}%
  \BibitemOpen
  \bibfield  {author} {\bibinfo {author} {\bibfnamefont {B.}~\bibnamefont
  {Bov\v{z}i\v{c}}}, \bibinfo {author} {\bibfnamefont {S.~L.}\ \bibnamefont
  {Das}},\ and\ \bibinfo {author} {\bibfnamefont {S.}~\bibnamefont {Svetina}},\
  }\href {https://doi.org/10.1039/C4SM02289K} {\bibfield  {journal} {\bibinfo
  {journal} {Soft Matter}\ }\textbf {\bibinfo {volume} {11}},\ \bibinfo {pages}
  {2479} (\bibinfo {year} {2015})}\BibitemShut {NoStop}%
\bibitem [{sup()}]{supplement}%
  \BibitemOpen
  \href@noop {} {}\bibinfo {note} {{See Supplemental Material at [URL will be
  inserted by publisher] for further details and an additional analysis of the
  model, which includes Refs.~\cite{Zhong1989, SoftMatterPhys, Deserno2015,
  Guckenberger2017, Goychuk2019}.}}\BibitemShut {Stop}%
\bibitem [{Note4()}]{Note4}%
  \BibitemOpen
  \bibinfo {note} {A further generalization yielding normal and tangential
  stresses involves variational surface calculus and is briefly outlined in the
  SM~\cite {supplement}. There, we show that the analysis presented here is
  valid in the limit of small deformations.}\BibitemShut {Stop}%
\bibitem [{\citenamefont {Kramers}(1940)}]{Kramers1940}%
  \BibitemOpen
  \bibfield  {author} {\bibinfo {author} {\bibfnamefont {H.~A.}\ \bibnamefont
  {Kramers}},\ }\href {https://doi.org/10.1016/S0031-8914(40)90098-2}
  {\bibfield  {journal} {\bibinfo  {journal} {Physica}\ }\textbf {\bibinfo
  {volume} {7}},\ \bibinfo {pages} {284} (\bibinfo {year} {1940})}\BibitemShut
  {NoStop}%
\bibitem [{\citenamefont {Bell}\ and\ \citenamefont
  {Terentjev}(2017)}]{Bell2017}%
  \BibitemOpen
  \bibfield  {author} {\bibinfo {author} {\bibfnamefont {S.}~\bibnamefont
  {Bell}}\ and\ \bibinfo {author} {\bibfnamefont {E.~M.}\ \bibnamefont
  {Terentjev}},\ }\href {https://doi.org/10.1016/j.bpj.2017.04.048} {\bibfield
  {journal} {\bibinfo  {journal} {Biophys. J.}\ }\textbf {\bibinfo {volume}
  {112}},\ \bibinfo {pages} {2439} (\bibinfo {year} {2017})}\BibitemShut
  {NoStop}%
\bibitem [{\citenamefont {Wu}\ \emph {et~al.}(2011)\citenamefont {Wu},
  \citenamefont {Park}, \citenamefont {Holyoak},\ and\ \citenamefont
  {Lutkenhaus}}]{Wu2011}%
  \BibitemOpen
  \bibfield  {author} {\bibinfo {author} {\bibfnamefont {W.}~\bibnamefont
  {Wu}}, \bibinfo {author} {\bibfnamefont {K.-T.}\ \bibnamefont {Park}},
  \bibinfo {author} {\bibfnamefont {T.}~\bibnamefont {Holyoak}},\ and\ \bibinfo
  {author} {\bibfnamefont {J.}~\bibnamefont {Lutkenhaus}},\ }\href
  {https://doi.org/10.1111/j.1365-2958.2010.07536.x} {\bibfield  {journal}
  {\bibinfo  {journal} {Mol. Microbiol.}\ }\textbf {\bibinfo {volume} {79}},\
  \bibinfo {pages} {1515} (\bibinfo {year} {2011})}\BibitemShut {NoStop}%
\bibitem [{\citenamefont {Ma}\ \emph {et~al.}(2017)\citenamefont {Ma},
  \citenamefont {Cai}, \citenamefont {Li}, \citenamefont {Jiao}, \citenamefont
  {Wu}, \citenamefont {O'Shaughnessy}, \citenamefont {De~Camilli},
  \citenamefont {Karatekin},\ and\ \citenamefont {Zhang}}]{Lu2017}%
  \BibitemOpen
  \bibfield  {author} {\bibinfo {author} {\bibfnamefont {L.}~\bibnamefont
  {Ma}}, \bibinfo {author} {\bibfnamefont {Y.}~\bibnamefont {Cai}}, \bibinfo
  {author} {\bibfnamefont {Y.}~\bibnamefont {Li}}, \bibinfo {author}
  {\bibfnamefont {J.}~\bibnamefont {Jiao}}, \bibinfo {author} {\bibfnamefont
  {Z.}~\bibnamefont {Wu}}, \bibinfo {author} {\bibfnamefont {B.}~\bibnamefont
  {O'Shaughnessy}}, \bibinfo {author} {\bibfnamefont {P.}~\bibnamefont
  {De~Camilli}}, \bibinfo {author} {\bibfnamefont {E.}~\bibnamefont
  {Karatekin}},\ and\ \bibinfo {author} {\bibfnamefont {Y.}~\bibnamefont
  {Zhang}},\ }\href {https://doi.org/10.7554/eLife.30493} {\bibfield  {journal}
  {\bibinfo  {journal} {eLife}\ }\textbf {\bibinfo {volume} {6}},\ \bibinfo
  {pages} {e30493} (\bibinfo {year} {2017})}\BibitemShut {NoStop}%
\bibitem [{\citenamefont {Dimova}(2014)}]{Dimova2014}%
  \BibitemOpen
  \bibfield  {author} {\bibinfo {author} {\bibfnamefont {R.}~\bibnamefont
  {Dimova}},\ }\href {https://doi.org/10.1016/j.cis.2014.03.003} {\bibfield
  {journal} {\bibinfo  {journal} {Adv. Colloid Interface Sci.}\ }\textbf
  {\bibinfo {volume} {208}},\ \bibinfo {pages} {225 } (\bibinfo {year}
  {2014})}\BibitemShut {NoStop}%
\bibitem [{\citenamefont {Nagle}\ \emph {et~al.}(2015)\citenamefont {Nagle},
  \citenamefont {Jablin}, \citenamefont {Tristram-Nagle},\ and\ \citenamefont
  {Akabori}}]{Nagle2015}%
  \BibitemOpen
  \bibfield  {author} {\bibinfo {author} {\bibfnamefont {J.~F.}\ \bibnamefont
  {Nagle}}, \bibinfo {author} {\bibfnamefont {M.~S.}\ \bibnamefont {Jablin}},
  \bibinfo {author} {\bibfnamefont {S.}~\bibnamefont {Tristram-Nagle}},\ and\
  \bibinfo {author} {\bibfnamefont {K.}~\bibnamefont {Akabori}},\ }\href
  {https://doi.org/10.1016/j.chemphyslip.2014.04.003} {\bibfield  {journal}
  {\bibinfo  {journal} {Chem. Phys. Lipids}\ }\textbf {\bibinfo {volume}
  {185}},\ \bibinfo {pages} {3 } (\bibinfo {year} {2015})}\BibitemShut
  {NoStop}%
\bibitem [{\citenamefont {Peleg}\ \emph {et~al.}(2011)\citenamefont {Peleg},
  \citenamefont {Disanza}, \citenamefont {Scita},\ and\ \citenamefont
  {Gov}}]{Peleg2011}%
  \BibitemOpen
  \bibfield  {author} {\bibinfo {author} {\bibfnamefont {B.}~\bibnamefont
  {Peleg}}, \bibinfo {author} {\bibfnamefont {A.}~\bibnamefont {Disanza}},
  \bibinfo {author} {\bibfnamefont {G.}~\bibnamefont {Scita}},\ and\ \bibinfo
  {author} {\bibfnamefont {N.}~\bibnamefont {Gov}},\ }\href
  {https://doi.org/10.1371/journal.pone.0018635} {\bibfield  {journal}
  {\bibinfo  {journal} {PLOS ONE}\ }\textbf {\bibinfo {volume} {6}},\ \bibinfo
  {pages} {1} (\bibinfo {year} {2011})}\BibitemShut {NoStop}%
\bibitem [{\citenamefont {Thalmeier}\ \emph {et~al.}(2016)\citenamefont
  {Thalmeier}, \citenamefont {Halatek},\ and\ \citenamefont
  {Frey}}]{Thalmeier2016}%
  \BibitemOpen
  \bibfield  {author} {\bibinfo {author} {\bibfnamefont {D.}~\bibnamefont
  {Thalmeier}}, \bibinfo {author} {\bibfnamefont {J.}~\bibnamefont {Halatek}},\
  and\ \bibinfo {author} {\bibfnamefont {E.}~\bibnamefont {Frey}},\ }\href
  {https://doi.org/10.1073/pnas.1515191113} {\bibfield  {journal} {\bibinfo
  {journal} {Proc. Natl. Acad. Sci. U.S.A.}\ ,\ \bibinfo {pages} {201515191}}
  (\bibinfo {year} {2016})}\BibitemShut {NoStop}%
\bibitem [{\citenamefont {Wu}\ \emph {et~al.}(2018)\citenamefont {Wu},
  \citenamefont {Su}, \citenamefont {Tong}, \citenamefont {Wu},\ and\
  \citenamefont {Liu}}]{Wu2018}%
  \BibitemOpen
  \bibfield  {author} {\bibinfo {author} {\bibfnamefont {Z.}~\bibnamefont
  {Wu}}, \bibinfo {author} {\bibfnamefont {M.}~\bibnamefont {Su}}, \bibinfo
  {author} {\bibfnamefont {C.}~\bibnamefont {Tong}}, \bibinfo {author}
  {\bibfnamefont {M.}~\bibnamefont {Wu}},\ and\ \bibinfo {author}
  {\bibfnamefont {J.}~\bibnamefont {Liu}},\ }\href
  {https://doi.org/10.1038/s41467-017-02469-1} {\bibfield  {journal} {\bibinfo
  {journal} {Nat. Commun.}\ }\textbf {\bibinfo {volume} {9}},\ \bibinfo {pages}
  {136} (\bibinfo {year} {2018})}\BibitemShut {NoStop}%
\bibitem [{\citenamefont {Huang}\ \emph {et~al.}(2003)\citenamefont {Huang},
  \citenamefont {Meir},\ and\ \citenamefont {Wingreen}}]{Huang2003}%
  \BibitemOpen
  \bibfield  {author} {\bibinfo {author} {\bibfnamefont {K.~C.}\ \bibnamefont
  {Huang}}, \bibinfo {author} {\bibfnamefont {Y.}~\bibnamefont {Meir}},\ and\
  \bibinfo {author} {\bibfnamefont {N.~S.}\ \bibnamefont {Wingreen}},\ }\href
  {https://doi.org/10.1073/pnas.2135445100} {\bibfield  {journal} {\bibinfo
  {journal} {Proc. Natl. Acad. Sci. U.S.A.}\ }\textbf {\bibinfo {volume}
  {100}},\ \bibinfo {pages} {12724} (\bibinfo {year} {2003})}\BibitemShut
  {NoStop}%
\bibitem [{\citenamefont {Halatek}\ and\ \citenamefont
  {Frey}(2012)}]{Halatek2012}%
  \BibitemOpen
  \bibfield  {author} {\bibinfo {author} {\bibfnamefont {J.}~\bibnamefont
  {Halatek}}\ and\ \bibinfo {author} {\bibfnamefont {E.}~\bibnamefont {Frey}},\
  }\href {https://doi.org/10.1016/j.celrep.2012.04.005} {\bibfield  {journal}
  {\bibinfo  {journal} {Cell Rep.}\ }\textbf {\bibinfo {volume} {1}},\ \bibinfo
  {pages} {741} (\bibinfo {year} {2012})}\BibitemShut {NoStop}%
\bibitem [{\citenamefont {Kl\"{u}nder}\ \emph {et~al.}(2013)\citenamefont
  {Kl\"{u}nder}, \citenamefont {Freisinger}, \citenamefont
  {Wedlich-S\"{o}ldner},\ and\ \citenamefont {Frey}}]{Klunder2013}%
  \BibitemOpen
  \bibfield  {author} {\bibinfo {author} {\bibfnamefont {B.}~\bibnamefont
  {Kl\"{u}nder}}, \bibinfo {author} {\bibfnamefont {T.}~\bibnamefont
  {Freisinger}}, \bibinfo {author} {\bibfnamefont {R.}~\bibnamefont
  {Wedlich-S\"{o}ldner}},\ and\ \bibinfo {author} {\bibfnamefont
  {E.}~\bibnamefont {Frey}},\ }\href
  {https://doi.org/10.1371/journal.pcbi.1003396} {\bibfield  {journal}
  {\bibinfo  {journal} {PLOS Comp. Biol.}\ }\textbf {\bibinfo {volume} {9}},\
  \bibinfo {pages} {1} (\bibinfo {year} {2013})}\BibitemShut {NoStop}%
\bibitem [{\citenamefont {Denk}\ \emph {et~al.}(2018)\citenamefont {Denk},
  \citenamefont {Kretschmer}, \citenamefont {Halatek}, \citenamefont {Hartl},
  \citenamefont {Schwille},\ and\ \citenamefont {Frey}}]{Denk2018}%
  \BibitemOpen
  \bibfield  {author} {\bibinfo {author} {\bibfnamefont {J.}~\bibnamefont
  {Denk}}, \bibinfo {author} {\bibfnamefont {S.}~\bibnamefont {Kretschmer}},
  \bibinfo {author} {\bibfnamefont {J.}~\bibnamefont {Halatek}}, \bibinfo
  {author} {\bibfnamefont {C.}~\bibnamefont {Hartl}}, \bibinfo {author}
  {\bibfnamefont {P.}~\bibnamefont {Schwille}},\ and\ \bibinfo {author}
  {\bibfnamefont {E.}~\bibnamefont {Frey}},\ }\href
  {https://doi.org/10.1073/pnas.1719801115} {\bibfield  {journal} {\bibinfo
  {journal} {Proc. Natl. Acad. Sci. U.S.A.}\ }\textbf {\bibinfo {volume}
  {115}},\ \bibinfo {pages} {4553} (\bibinfo {year} {2018})}\BibitemShut
  {NoStop}%
\bibitem [{\citenamefont {Goryachev}\ and\ \citenamefont
  {Leda}(2017)}]{Goryachev2017}%
  \BibitemOpen
  \bibfield  {author} {\bibinfo {author} {\bibfnamefont {A.~B.}\ \bibnamefont
  {Goryachev}}\ and\ \bibinfo {author} {\bibfnamefont {M.}~\bibnamefont
  {Leda}},\ }\href {https://doi.org/10.1091/mbc.e16-10-0739} {\bibfield
  {journal} {\bibinfo  {journal} {Mol. Biol. Cell}\ }\textbf {\bibinfo {volume}
  {28}},\ \bibinfo {pages} {370} (\bibinfo {year} {2017})}\BibitemShut
  {NoStop}%
\bibitem [{\citenamefont {Halatek}\ and\ \citenamefont
  {Frey}(2018)}]{Halatek2018b}%
  \BibitemOpen
  \bibfield  {author} {\bibinfo {author} {\bibfnamefont {J.}~\bibnamefont
  {Halatek}}\ and\ \bibinfo {author} {\bibfnamefont {E.}~\bibnamefont {Frey}},\
  }\href {https://doi.org/10.1038/s41567-017-0040-5} {\bibfield  {journal}
  {\bibinfo  {journal} {Nat. Phys.}\ }\textbf {\bibinfo {volume} {14}},\
  \bibinfo {pages} {507} (\bibinfo {year} {2018})}\BibitemShut {NoStop}%
\bibitem [{\citenamefont {Zhong-Can}\ and\ \citenamefont
  {Helfrich}(1989)}]{Zhong1989}%
  \BibitemOpen
  \bibfield  {author} {\bibinfo {author} {\bibfnamefont {O.-Y.}\ \bibnamefont
  {Zhong-Can}}\ and\ \bibinfo {author} {\bibfnamefont {W.}~\bibnamefont
  {Helfrich}},\ }\href {https://doi.org/10.1103/PhysRevA.39.5280} {\bibfield
  {journal} {\bibinfo  {journal} {Phys. Rev. A}\ }\textbf {\bibinfo {volume}
  {39}},\ \bibinfo {pages} {5280} (\bibinfo {year} {1989})}\BibitemShut
  {NoStop}%
\bibitem [{\citenamefont {Doi}(2013)}]{SoftMatterPhys}%
  \BibitemOpen
  \bibfield  {author} {\bibinfo {author} {\bibfnamefont {M.}~\bibnamefont
  {Doi}},\ }\href@noop {} {\emph {\bibinfo {title} {Soft Matter Physics}}}\
  (\bibinfo  {publisher} {Oxford University Press},\ \bibinfo {year}
  {2013})\BibitemShut {NoStop}%
\bibitem [{\citenamefont {Deserno}(2015)}]{Deserno2015}%
  \BibitemOpen
  \bibfield  {author} {\bibinfo {author} {\bibfnamefont {M.}~\bibnamefont
  {Deserno}},\ }\href {https://doi.org/10.1016/j.chemphyslip.2014.05.001}
  {\bibfield  {journal} {\bibinfo  {journal} {Chem. Phys. Lipids}\ }\textbf
  {\bibinfo {volume} {185}},\ \bibinfo {pages} {11 } (\bibinfo {year}
  {2015})}\BibitemShut {NoStop}%
\bibitem [{\citenamefont {Guckenberger}\ and\ \citenamefont
  {Gekle}(2017)}]{Guckenberger2017}%
  \BibitemOpen
  \bibfield  {author} {\bibinfo {author} {\bibfnamefont {A.}~\bibnamefont
  {Guckenberger}}\ and\ \bibinfo {author} {\bibfnamefont {S.}~\bibnamefont
  {Gekle}},\ }\href {https://doi.org/10.1088/1361-648x/aa6313} {\bibfield
  {journal} {\bibinfo  {journal} {J. Phys. Condens. Matter}\ }\textbf {\bibinfo
  {volume} {29}},\ \bibinfo {pages} {203001} (\bibinfo {year}
  {2017})}\BibitemShut {NoStop}%
\bibitem [{\citenamefont {Goychuk}\ \emph {et~al.}()\citenamefont {Goychuk},
  \citenamefont {Hassan}, \citenamefont {Wigbers},\ and\ \citenamefont
  {Frey}}]{Goychuk2019}%
  \BibitemOpen
  \bibfield  {author} {\bibinfo {author} {\bibfnamefont {A.}~\bibnamefont
  {Goychuk}}, \bibinfo {author} {\bibfnamefont {L.}~\bibnamefont {Hassan}},
  \bibinfo {author} {\bibfnamefont {M.}~\bibnamefont {Wigbers}},\ and\ \bibinfo
  {author} {\bibfnamefont {E.}~\bibnamefont {Frey}},\ }\href@noop {} {\bibinfo
  {journal} {in preparation}\ }\BibitemShut {NoStop}%
\end{thebibliography}
\end{document}


\title{Mechanical cooperativity of protein-membrane binding -- Supplementary Material}
\author{Andriy Goychuk}
\affiliation{Arnold Sommerfeld Center for Theoretical Physics and Center for NanoScience, Department of Physics, Ludwig-Maximilians-Universit\"at M\"unchen, Theresienstr. 37, D-80333 Munich, Germany}
\author{Erwin Frey}
\affiliation{Arnold Sommerfeld Center for Theoretical Physics and Center for NanoScience, Department of Physics, Ludwig-Maximilians-Universit\"at M\"unchen, Theresienstr. 37, D-80333 Munich, Germany}
\date{\today}

\maketitle


\section{Equivalent forms of the free energy density}
%


In the literature one finds various equivalent representations of the free energy density describing the thermodynamics of membrane-bound proteins that are coupled to membrane curvature. 
This section serves to show how these can be related to our model.   
As described in the main text, the free energy density is given by
\begin{multline}
	f (u,m)
	=
	\tfrac12 \kappa \, (u \,{-}\, u_0)^2 \\
	+ m \, 
	\bigl[
	E_\text{opt} \,(1+\gamma \, m^6)
	+
	\tfrac{1}{2} \, \epsilon \, 
	(u - u_\text{opt})^2
	\bigr] 
	\, .
\label{eq:oldformulation}
\end{multline}
Upon collecting all terms that depend on the membrane deformation $u$, an equivalent form of the free energy is:
\begin{multline}
	f (u,\widetilde{m})
	=
	\tfrac12 \kappa \, (1+\widetilde{m}) \, (u - u_\times)^2 \\
	+ \left[
	\frac{\widetilde{m}}{1+\widetilde{m}} 
	+\widetilde{m} \bigl(
	1 + \widetilde{\gamma} \, \widetilde{m}^6
	\bigr) \,
	\frac{E_\text{opt}}{\Delta E}
	\right] \, \Delta f \, .
\label{eq:reformulation}
\end{multline}
Here, we have defined a characteristic membrane protein density, $m_\times \,{=}\, \kappa / \epsilon$, and rescaled our variables accordingly, $\widetilde{m} \,{:=}\,  m/m_\times$ and $\widetilde{\gamma} \,{:=}\, \gamma / m_\times^6$.
In this form of the free energy, the characteristic membrane conformation, $u_\times \,{=}\, u_0 \,{+}\, (u_\text{opt} \,{-}\, u_0) \, {\widetilde{m}}/(1 {+}\, \widetilde{m})$, directly gives the membrane deformation minimizing the free energy. 
Moreover, one can directly read off that protein attachment enhances the stiffness parameter: $\kappa_\text{eff}=\kappa \, (1+\widetilde{m})$.
The second line of Eq.~\eqref{eq:reformulation} represents a free energy density contribution that only depends on the protein density but not on the membrane conformation. 
This corresponds to Eq.~\eqref{main-eq::free_energy_simple} in the main text, as $u=u_\times$ minimizes Eq.~\eqref{eq:reformulation}.

Next we show how one can use Eq.~\eqref{eq:reformulation} to obtain a Helfrich bending energy containing a spontaneous curvature that linearly depends on the membrane protein density~\cite{Shlomovitz2009,Zhu2012}. 
We set $u\equiv H$ and $u_\text{opt} \equiv H_\text{opt}$, and assume that the membrane is symmetric in the absence of membrane-bound proteins.
Then, the intrinsic spontaneous curvature of the membrane vanishes, $u_0 \equiv H_0=0$.
If the density of membrane-bound proteins is sufficiently small, $\widetilde{m}\ll 1$, then the first line of Eq.~\eqref{eq:reformulation} simplifies to:
\begin{equation}
	\tfrac12 \kappa \, (H - \widetilde{m} \, H_\text{opt})^2 \, .
\label{eq:helfrich_our}
\end{equation}
This term can be rewritten in terms of protein surface coverage, $\theta = m/m_\text{s} = \widetilde{m}/\widetilde{m}_\text{s}$, where $m_\text{s}$ is the  surface saturation density of proteins.
We also rescale the optimal curvature, $H_{\text{opt}/\theta} = \widetilde{m}_\text{s} \, H_\text{opt}$, to arrive at:
\begin{equation}
	\tfrac12 \kappa \, (H - \theta \, H_{\text{opt}/\theta})^2 \, .
\label{eq:helfrich_common}
\end{equation}
The above expression is sometimes used when coupling proteins to membrane curvature~\cite{Shlomovitz2009,Zhu2012}.

%


\section{Additional free energy contributions}

In the main text, we have restricted ourselves to repulsive interactions between  the proteins, $f_\text{rep}=\gamma m^6$, mainly in order to bound the surface density of surface proteins and to introduce a saturation coverage.
Here, we study the additional effects of entropic mixing of proteins and attractive interactions between proteins.
The main conclusion drawn from the following analysis is: 
\begin{enumerate*}
	\item Importantly, in the high protein specificity and low protein density regime, where we find protein recruitment in the main text, both of these contributions have only minor effects on the binding kinetics. 
	\item In the low protein specificity and high protein density regime, both of these contributions play major roles, as we will discuss below. 
\end{enumerate*} 

\subsection{Mixing entropy of membrane-bound proteins}
\label{sec:mixing_entropy}

\begin{figure}[bt]
\includegraphics{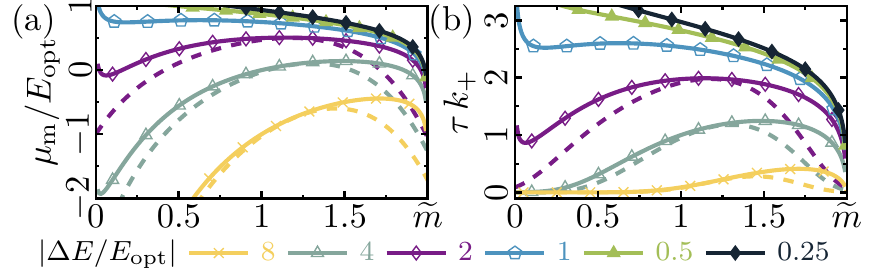}
\caption{
Comparison between the influence of mixing entropy (solid lines), and explicit repulsive interactions between proteins (dashed lines) on binding kinetics.
Membrane chemical potential (a),  $\mu_\text{m}/E_\text{opt}$, and kinetic attachment rate (b), $k_+\, \tau$, plotted as a function of the density of membrane-bound proteins, $m/m_\times$, for a series of different protein binding specificities, $|\Delta E/E_\text{\normalfont  opt}|$, indicated in the graph.
The optimal binding energy is given by $E_\text{opt} \,{=}\, {-}5 k_\text{\normalfont B} T$, and the membrane saturation concentration by $\widetilde{m}_\text{s}=2.0$.
}
\label{fig:coop_entropic}
\end{figure}

The free energy density including the mixing entropy but neglecting repulsive interactions is given by
\begin{multline}
	f (u,m)
	=
	\tfrac12 \kappa \, (u \,{-}\, u_0)^2
	+ m \, 
	\bigl[
	E_\text{opt} \,
	+
	\tfrac{1}{2} \, \epsilon \, 
	(u - u_\text{opt})^2
	\bigr] \\
	+ k_\text{B} T \left[m \, \ln\left(\tfrac{m}{m_\text{s}}\right) + (m_\text{s} - m)\,\ln\left(\tfrac{m_\text{s}-m}{m_\text{s}}\right)\right]
	\, ,
\end{multline}
where the second line denotes the mixing entropy contribution, with $m_\text{s}$ as the saturation density of the membrane.
We proceed similar as described in the main text by adiabatically eliminating the mechanical degrees of freedom.
Then, we find the following expression for the chemical potential at the membrane:
\begin{equation}
	\frac{\mu_\text{m} (\widetilde{m})}
	     {E_\text{opt}} 
	= 
	1 + 
	\frac{\Delta E}{E_\text{opt}} \,
	\frac{1}{(1 + \widetilde{m})^2} +
	\frac{k_\text{B} T}{E_\text{opt}}\ln\left[\frac{\widetilde{m}}{\widetilde{m}_\text{s}-\widetilde{m}}\right]
	\, ,
\label{eq:chempot_mixing}
\end{equation}
where $\widetilde{m}_\text{s} \coloneqq m_0/m_\times$ is the non-dimensionalized saturation density.
Comparing the chemical potentials, which include contributions from
\begin{enumerate*}[label={(\Alph*)}]
	\item mixing entropy, Eq.~\eqref{eq:chempot_mixing}, \emph{or}
	\item repulsion between proteins, Eq.~\eqref{main-eq::binding_energy_total} in the main text,
\end{enumerate*} 
we find that both variants show strong repulsion at high protein densities [Fig.~\ref{fig:coop_entropic}a].
At low protein densities, we find that both chemical potential variants become more similar as we increase protein specificity, thereby reducing the relative weight of the entropic/repulsive terms in Eqs.~\eqref{eq:chempot_mixing}~and~\eqref{main-eq::binding_energy_total}, respectively [Fig.~\ref{fig:coop_entropic}, compare solid with dashed lines].


We proceed as described in the main text and use the chemical potential at the membrane to determine the protein binding rates.
Analogously to the main text, the protein attachment rates increase nonlinearly as a function of membrane-bound protein density, for high protein specificities and low protein densities [Fig.~\ref{fig:coop_entropic}b, crosses and empty triangles].
As discussed above, in this regime, where recruitment can be observed, the mixing entropy does not play a major role for the binding kinetics.
However, for low protein specificities, the binding rates monotonically decrease (for all densities) with increasing membrane-bound protein density [Fig.~\ref{fig:coop_entropic}b, filled markers], because the proteins always compete for the available space.
Note that, in contrast, in the main text, the attachment rates only decrease when  repulsive interactions between proteins become dominant at high protein densities [Fig.~\ref{main-fig::free_energy}b].


\subsection{Attractive interactions between proteins}
%

\begin{figure}[bt]
\includegraphics{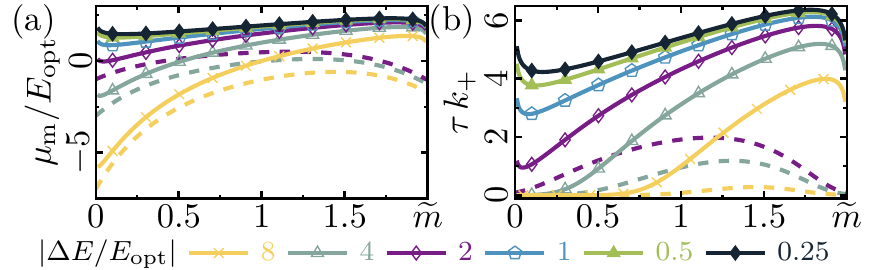}
\caption{
Comparison between the influence of mixing entropy with explicit attractive interactions between proteins (solid lines), and explicit repulsive interactions between proteins (dashed lines) on binding kinetics.
Membrane chemical potential (a),  $\mu_\text{m}/E_\text{opt}$, and kinetic attachment rate (b), $k_+\, \tau$, plotted as a function of the density of membrane-bound proteins, $m/m_\times$, for a series of different protein binding specificities, $|\Delta E/E_\text{\normalfont  opt}|$, indicated in the graph.
The optimal binding energy is given by $E_\text{opt} \,{=}\, {-}5 k_\text{\normalfont B} T$, the membrane saturation concentration by $\widetilde{m}_\text{s}=2.0$, and the attractive interaction strength between proteins is given by $\widetilde{\chi}=2.5k_\text{B}T$.
}
\label{fig:coop_entropic_attractive}
\end{figure}

Motivated by Ref.~\cite{Zhu2012}, we additionally study the effect of attractive interactions between membrane-bound proteins, which could for example arise from a conformational change upon attachment to the membrane.
Then, the free energy density is captured by a (modified) Flory-Huggins theory, including interactions between proteins and the membrane:
\begin{multline}
f (u,m)
=
\tfrac12 \kappa \, (u \,{-}\, u_0)^2
+ m \, 
\bigl[
E_\text{opt} \,
+
\tfrac{1}{2} \, \epsilon \, 
(u - u_\text{opt})^2
\bigr]
\\
+ k_\text{B} T \left[m \, \ln\left(\tfrac{m}{m_\text{s}}\right) + (m_\text{s}-m)\,\ln\left(\tfrac{m_\text{s}-m}{m_\text{s}}\right)\right]
- \chi m^2
\, ,
\end{multline}
where $m_\text{s}$ is the saturation density of the membrane and $\chi$ encodes the strength of the attractive interactions between proteins.
We proceed similar as described in the main text by adiabatically eliminating the mechanical degrees of freedom.
Then, we find the following expression for the chemical potential at the membrane:
\begin{multline}
	\frac{\mu_\text{m} (m)}
	     {E_\text{opt}} 
	= 
	1 + 
	\frac{\Delta E}{E_\text{opt}} \,
	\frac{1}{(1 + \widetilde{m})^2} \\
	+ \frac{k_\text{B} T}{E_\text{opt}}\ln\left[\frac{\widetilde{m}}{\widetilde{m}_\text{s}-\widetilde{m}}\right]
	- \frac{2 \,\widetilde{\chi}}{E_\text{opt}}\widetilde{m}
	\, .
\label{eq:chempot_mixing_attractive}
\end{multline}
where $\widetilde{m}_\text{s} \coloneqq m_0/m_\times$ is the non-dimensionalized saturation density and $\widetilde{\chi} \coloneqq \chi \, m_\times$ is the rescaled protein-protein interaction strength.
For small membrane-bound protein densities, the contribution of attractive interactions between proteins to the chemical potential vanishes, as the last term in Eq.~\eqref{eq:chempot_mixing_attractive} becomes negligible.
Then, one obtains a similar behavior as in Sec.~\ref{sec:mixing_entropy} [Fig.~\ref{fig:coop_entropic_attractive}a, compare solid with dashed lines]: with increasing protein specificity, the chemical potential, Eq.~\eqref{eq:chempot_mixing_attractive}, approaches the values from the main text, given by Eq.~\eqref{main-eq::binding_energy_total}.

We proceed as described in the main text and use the chemical potential at the membrane to determine the protein binding rates.
Analogously to the main text, for high protein specificities, we find a nonlinear recruitment of proteins from the cytosol to the membrane [Fig.~\ref{fig:coop_entropic_attractive}b, crosses and empty triangles].
At low protein densities, the contribution from the attractive interactions between proteins becomes negligible.
Then, recruitment originates from the mechanochemical interactions between proteins and membrane, as described in detail in the main text.
At high protein densities, where mechanochemical contributions to the binding rates saturate regardless of protein specificity, attractive interactions between proteins becomes dominant and lead to recruitment.
Therefore, we conclude that cooperative protein recruitment from the cytosol to the membrane is obtained for the following cases
\begin{enumerate*}[label=(\roman*)]
	\item at low protein densities by a highly specific mechanochemical coupling between proteins and membrane (tunable nonlinear recruitment), \emph{or}
	\item at high protein densities by attractive chemical interactions between proteins. Here, however, we observe only a linear/sublinear increase of the attachment rates with the membrane-bound protein density [Fig.~\ref{fig:coop_entropic_attractive}b, empty diamonds and pentagons, filled triangles and diamonds].
\end{enumerate*}

\section{Protein binding through anchor insertion}
%


In the main text, we have constructed a description of protein-membrane interactions via symmetry arguments.
Then, we have argued that it applies to different types of membrane deformations, like lateral strain by lipid density changes.
However, one might ask how single proteins can affect the local lipid density, given that any changes of the lipid density \emph{in between} membrane-targeting protein anchors should relax quickly.
While this is certainly correct, at a finite density of membrane-bound proteins, the free energy density is affected by three factors:
\begin{enumerate*}[label=(\roman*)]
	\item lipid-lipid interactions,
	\item protein-lipid interactions, and
	\item the corresponding mixing entropy.
\end{enumerate*}
In the following, we derive such free energy, and thereby provide a more detailed rationalization for the description of protein-lipid coupling.

\begin{figure}[tb]
	\includegraphics{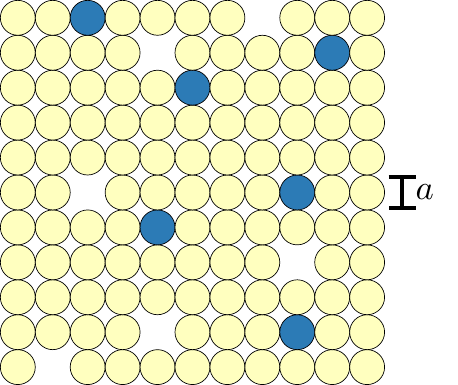}
	\caption{
	Lattice gas model of a membrane with $N$ lattice sites of size $a^2$.
	Most of the available sites are occupied by $N_\text{l}$ phospholipids (yellow), which exhibit strong attractive interactions with other phospholipids.
	The remaining sites are either occupied by $N_\text{p}$ protein anchors (blue), which bind to the membrane through attractive interactions with phospholipids, or remain empty.
	}
	\label{fig::latticegas}
\end{figure}

We assume that the membrane can be described as a ternary lattice gas consisting of $N_\text{l}$ lipids, $N_\text{p}$ protein anchors, and $N-N_\text{l} - N_\text{p}$ unoccupied lattice sites [Fig.~\ref{fig::latticegas}].
In a lipid membrane, there should be much more lipids than protein anchors or unoccupied lattice sites: $N_\text{l}\gg N_\text{p}$ and $N_\text{l}\gg N-N_\text{l} - N_\text{p}$.
To make our calculations as simple as possible, we assume that each lattice site (size of $a^2$) can be occupied by one of these three key players.
Then, the mixing entropy contribution to the free energy density of such a ternary mixture is given by~\cite{SoftMatterPhys}:
\begin{multline}
	f_\text{mix}=
	k_\text{B}T
	\Big[
	\rho \ln\left(\tfrac{\rho}{\rho_\text{s}}\right)+ 
	m \ln\Big(\tfrac{m}{\rho_\text{s}}\Big) \\
	+ (\rho_\text{s}-\rho-m) \ln\Big(\tfrac{\rho_\text{s}-\rho-m}{\rho_\text{s}}\Big)
	\Big],
\label{eq:mixing}
\end{multline}
where we have introduced the saturation density $\rho_\text{s} = 1/a^2$, the lipid density $\rho = N_\text{l}/(N a^2)$ and the protein density $m = N_\text{p}/(N a^2)$.
In addition to mixing entropy, we assume that lipids strongly attract each other with an interaction energy $E_\text{ll}\gg k_\text{B}T$.
Furthermore, protein anchors and lipids also mutually attract each other $E_\text{lp} > k_\text{B}T$; this attraction should exceed thermal energy to make protein binding favorable.
In summary, we then obtain the following Flory-Huggins free energy:
\begin{equation}
	f=
	- \tfrac{E_\text{\vphantom{p}ll}}{\rho_\text{s}} \, \rho^2
	- \tfrac{E_\text{lp}}{\rho_\text{s}} \, \rho \, m
	+ f_\text{mix} \, .
\end{equation}
We collect the terms in the mixing free energy, Eq.~\eqref{eq:mixing}, into a protein mixing free energy density (this contribution does not depend on the lipid density),
\begin{subequations} 
\begin{equation}
	f_\text{mix/m}=
	k_\text{B}T
	\Big[
		m \ln\Big(\tfrac{m}{\rho_\text{s}}\Big) +
		(\rho_\text{s}-m) \ln\Big(\tfrac{\rho_\text{s}-m}{\rho_\text{s}}\Big)
	\Big] \, ,
\end{equation}
and the remainder, $f_\text{mix/r} = f_\text{mix} - f_\text{mix/m}$,
\begin{equation}
	f_\text{mix/r}=
	k_\text{B}T
	\Big[
	\rho\ln\left(\tfrac{\rho}{\rho_\text{s}-\rho-m}\right) +
	(\rho_\text{s}-m) \ln\Big(\tfrac{\rho_\text{s}-\rho-m}{\rho_\text{s}-m}\Big)
	\Big] \, .
\end{equation}
Then, given that membrane-bound protein densities should be small, we expand the remainder $f_\text{mix/r}$ into a Taylor series, to first order in $m$ in the vicinity of $m=0$:
\begin{equation}
	f_\text{mix/r}= 
	f_{\text{mix}/\rho}
	-
	m \,
	k_\text{B}T
	\ln\big(
	\tfrac{\rho_s-\rho}{\rho_s}
	\big) \, ,
\end{equation}
where we have introduced the lipid mixing free energy density
\begin{equation}
	f_{\text{mix}/\rho} = 
	k_\text{B}T
	\Big[
	\rho\ln\left(\tfrac{\rho}{\rho_\text{s}}\right) +
	(\rho_s-\rho)\ln\left(\tfrac{\rho_s-\rho}{\rho_\text{s}}\right)
	\Big] \, .
\end{equation}
\end{subequations}
In summary, for small membrane-bound protein densities, the free energy density is given by:
\begin{multline}
	f = 
	- \tfrac{E_\text{\vphantom{p}ll}}{\rho_\text{s}} \, \rho^2
	+ f_{\text{mix}/\rho} \\
	+ 
	f_\text{mix/m}
	-
	m 
	\left[
	\tfrac{E_\text{lp}}{\rho_\text{s}} \, \rho
	+
	k_\text{B}T
	\ln\left(
	\tfrac{\rho_s-\rho}{\rho_s}
	\right)
	\right] \, .
\label{eq:free_energy_approx}
\end{multline}
For large attractive interaction energies between lipids, the first line of Eq.~\eqref{eq:free_energy_approx} will have a minimum at an intrinsically preferred lipid density $\rho_0$, which is close to the saturation density.
Then, we can expand the first line of Eq.~\eqref{eq:free_energy_approx} to second order in the vicinity of its minimum:
\begin{equation}
	-\tfrac{E_\text{\vphantom{p}ll}}{\rho_\text{s}} \, \rho^2 + f_{\text{mix}/\rho}
	\approx 
	\tfrac12\kappa(\rho-\rho_0)^2 + \text{const} \, .
\end{equation}
In addition, we also expand the last term of Eq.~\eqref{eq:free_energy_approx} to second order in $\rho$ in the vicinity of its minimum 
\begin{equation}
	\rho_\text{opt}=\rho_\text{s} \left(1-\tfrac{k_\text{B}T}{E_\text{lp}}\right) \, ,
\end{equation}
where $E_\text{lp}>k_\text{B}T$.
Having done all that, we finally arrive at an expression which is analogous to the one used in the main text (plus an expression that contains the mixing entropy of proteins),
\begin{equation}
	f=
	\tfrac12\kappa(\rho-\rho_0)^2
	+
	m \big[
	E_\text{opt}
	+\tfrac12 \epsilon \big(\rho-\rho_\text{opt}\big)^2
	\big]
	+ f_{\text{mix}/m} \, ,
\end{equation}
where the optimal binding energy is given by
\begin{equation}
	E_\text{opt}=k_\text{B}T\Big(1 + \ln\tfrac{E_\text{lp}}{k_\text{B}T} - \tfrac{E_\text{lp}}{k_\text{B}T}\Big) \, ,
\end{equation}
and the coupling between proteins and membrane is given by
\begin{equation}
	\epsilon = \frac{2 E_\text{lp}^2}{\rho_\text{s}^2 \, k_\text{B}T} \, .
\end{equation}
Therefore, an explicit thermodynamic calculation yields the same results that were obtained in the main text through symmetry arguments.
Effectively, this models a situation where high lipid densities prevent anchor insertion by steric repulsion, while low lipid densities lack the attractive interaction between the lipids and the anchor that is necessary for binding.
Then, protein anchors, which are inserted into the inner membrane leaflet, induce bulk strain in the inner leaflet by increasing the density of lipids, while leaving the outer leaflet unperturbed.

\section{Realistic chemical potential profile}

In the main text, we have considered the special case where a protein interacts with the membrane across the whole reaction compartment ($\xi=d$). 
Here we also discuss the general case $\xi>d$, where a protein diffuses freely (flat chemical potential) in the region $x \in [d\dots\xi]$, and interacts (ramp potential) with the membrane within the range $x\in[0 \dots d]$.
Then, the kinetic rates for biding and unbinding are given by:
\begin{eqnarray}
	k_\pm \, \tau &=& \xi^2 \left((\xi-d)^2 + 2 \,  d^2 \Gamma_2 \mp 2 d (\xi-d) \Gamma_1 \right)^{-1} \, ,\\
	\Gamma_2 &=& \left(\frac{k_\text{B} T}{\mu_\text{m}}\right)^2 \left(\text{e}^{\pm \mu_\text{m}/k_\text{B} T} \mp \frac{\mu_\text{m}}{k_\text{B} T} -1\right) \, , \\ 
	\Gamma_1 &=& \left(\frac{k_\text{B} T}{\mu_\text{m}}\right) \left(\text{e}^{\pm \mu_\text{m}/k_\text{B} T} -1 \right) \, ,
\end{eqnarray}
where $\mu_\text{m}$ is the chemical potential of membrane-bound proteins and $\tau = \xi^2 / 2D$ is the basic timescale of diffusion across the compartment.
\begin{figure}[b]
	\includegraphics{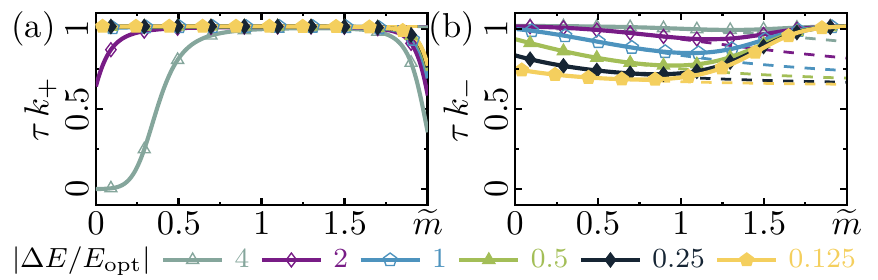}
	\caption{
	Membrane attachment and detachment rates of proteins for $E_\text{\normalfont  opt} = -5 k_\text{B} T$ and $\xi - d = 100 \,d$.
	Solid lines represent $\widetilde{\gamma}=-0.004$; dashed lines represent $\widetilde{\gamma}=0$.
	(a) The attachment rate increases as a function of the concentration $\widetilde{m}$ of membrane-bound proteins until the membrane saturates.
	(b) The detachment rate remains more or less constant as a function of the concentration $\widetilde{m}$ of membrane-bound proteins.
	}
	\label{fig::binding_cooperativity}
\end{figure}
For $\xi\rightarrow d$ we recover the case discussed in the main text [Eq.~\eqref{main-eq::rates}], while for $d\rightarrow 0$ there is no spatial variation in the chemical potential and the resulting kinetic rates reduce to $k_\pm = \tau^{-1}$.

We will discuss a situation where the protein does not interact with the membrane throughout most of the reaction compartment ($\xi - d \gg d$); for specificity, we consider $\xi - d = 100 \,d$.
Then, we observe a much smaller variation of the detachment rates compared to the main text, where $\xi - d = 0$ [compare Fig.~\ref{fig::binding_cooperativity}b~with~Fig.~\ref{main-fig::binding_cooperativity}b].
This reduction can be explained as follows. 
After detachment the protein starts diffusing in the chemical potential landscape at the membrane. 
Therefore, to leave the reaction compartment, the protein first has to traverse the interaction range $d$, across which it interacts with the membrane and senses a steep chemical potential [Fig.~\ref{main-fig::illustration}a, bottom panel].
The shorter the interaction range $d$, the faster the protein leaves the steep chemical potential and enters the region of free diffusion, which is why for $\xi - d \gg d$ the rate-limiting timescale given by time $\tau$ that it takes to diffuse across the reaction compartment.

In contrast, as in the main text, the attachment rate is a highly nonlinear function of the membrane-bound protein concentration $\widetilde{m}$ [Fig.~\ref{fig::binding_cooperativity}a].
This is surprising since one might expect that, similar as for the detachment process, diffusion in the extended flat region of the chemical potential is the dominant rate-limiting factor.
However, this is not the case. 
To understand the origin for this difference, note that the boundary conditions of these two processes are different.
During unbinding, the protein detaches from the membrane, whose reflective boundary condition quickly drives the protein across the interaction range $d$.
The situation is genuinely different for the attachment process, where the protein originates from the bulk and first has to diffuse across the distance $\xi -d$ to reach the membrane.
Then, near the membrane, the protein enters the interaction range $d$ with its steep chemical potential, which is repulsive for small membrane-bound protein concentrations and only becomes attractive for high protein concentrations.
As there is no nearby reflective boundary (effectively driving the protein away from it), the rate limiting factor of the attachment process becomes the time needed to diffuse against the steep chemical potential and to reach the absorbing boundary at the membrane.
\begin{figure}[tb]
	\includegraphics{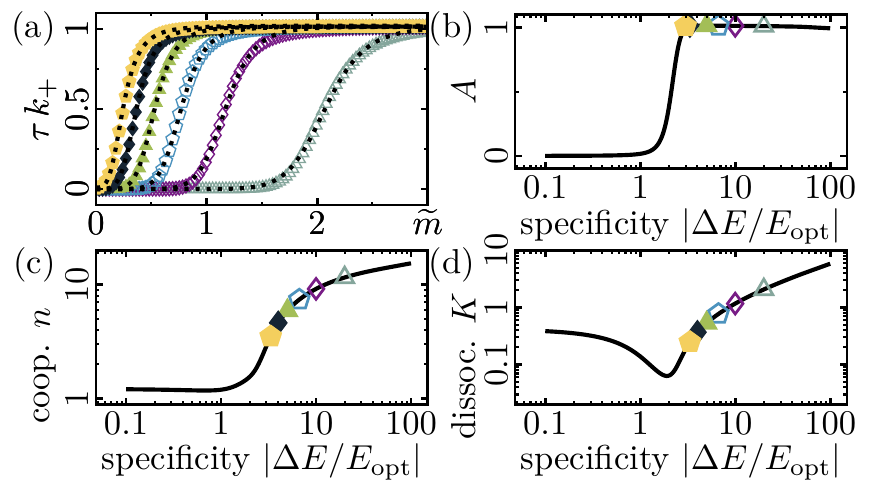}
	\caption{
	Effective representation of binding cooperativity for the parameter values in Fig.~\ref{fig::binding_cooperativity}.
	(a) The fit $\tau \, (k_+ \,-\, k_+|_{\widetilde{m}=0})  \approx A \, \widetilde{m}^n / (K^n + \widetilde{m}^n)$ (indicated by the dotted lines) is in good agreement with the attachment rate.	
	(b) The scale factor $A$, which indicates whether the attachment rate is constant or follows a Hill-like curve, indicates a transition from a diffusion-limited to a highly nonlinear attachment process at a protein specificity $|\Delta E / E_\text{opt}| \approx 2$.
	Both the cooperativity coefficient (c) and the apparent dissociation constant (d) increase with the specificity of the protein $|\Delta E / E_\text{opt}|$. 
	Note that both measures only capture the prominent properties of the attachment process when $A\approx 1$.
	}
	\label{fig::effective}
\end{figure}

As discussed above, the detachment rate only varies slightly depending on the membrane-bound protein concentration $\tau \, k_- \in [0.7 \dots 1]$.
The nonlinear dependence of the binding rate $k_+$ on the membrane protein density $\widetilde{m}$ is well approximated by a Hill-like curve: $\tau \, (k_+ \,-\, k_+|_{\widetilde{m}=0})  \approx A \, \widetilde{m}^n / (K^n + \widetilde{m}^n)$ [Fig.~\ref{fig::effective}a], with $A\approx 1$ [Fig.~\ref{fig::effective}b].
Similar as in the main text [cf. Fig.~\ref{main-fig::binding_cooperativity}d], this shows that proteins with a higher specificity $|\Delta E \,/\, E_\text{opt}|$ show higher cooperativity $n$ [Fig.~\ref{fig::effective}c].
Here, the scale factor $A$ indicates whether the attachment rate is constant or nonlinear: $k_+ \approx k_+|_{\widetilde{m}=0}$ if $A=0$, while $k_+ \approx \tau^{-1} \, \widetilde{m}^n / (K^n + \widetilde{m}^n)$ if $A\approx 1$; note that $\text{\normalfont max}(k_+) \approx \tau^{-1}$ [cf. Figs.~\ref{fig::binding_cooperativity}a,~\ref{fig::effective}a].
We find that the attachment process becomes highly non-linear above a protein specificity $|\Delta E / E_\text{opt}| \approx 2$.
Then, both the cooperativity coefficient [Fig.~\ref{fig::effective}c] and the apparent dissociation constant [Fig.~\ref{fig::effective}d] increase with the specificity of the protein.

Taken together, as in the main text, increasing protein density deforms the membrane towards a binding-favorable conformation, and ultimately increases the attachment rate in a pronounced nonlinear fashion.

\section{Variational treatment of the membrane}

In this section, we sketch how one can extend the model presented in the main text to describe membranes of arbitrary shape.
Furthermore, we show how gradients in membrane conformation drive in-plane flows of membrane-bound proteins.

\subsection{Shape equation for arbitrary deformations}
\label{sec:arbitrary_deformations}

In the main text, we have adiabatically eliminated the mechanical degrees of freedom in the free energy by finding the membrane conformation that \emph{locally} minimizes the free energy density.
However, it is a priori not clear why this should also correspond to the minimum of the \emph{global} free energy functional.
In this section, we show that the results obtained from the main text are valid as long as membrane deformations are sufficiently small.

We assume that proteins couple to the membrane curvature.
Then, analogously to the main text, the free energy density is given by:
\begin{multline}
	f (u,m)
	=
	\tfrac12 \kappa \, (H \,{-}\, H_0)^2\\
	+ m \, 
	\bigl[
	E_\text{opt}
	+
	\tfrac{1}{2} \, \epsilon \, 
	(H - H_\text{opt})^2
	\bigr]
	+ f_{\text{PP}}(m)
	\, ,
\label{eq:free_energy}
\end{multline}
where $f_{\text{PP}}(m)$ models direct interactions between proteins (which can be attractive, repulsive, or entropic).
Note that, in principle, one could also consider proteins that couple to the lipid density in the membrane.
Such a coupling could be achieved by insertion of lipid-targeting anchors into the inner leaflet of the membrane, while leaving the outer leaflet unperturbed.
To account for such a coupling, one would separately consider the lipid densities in both membrane leaflets, which leads to a description of membrane deformations within the area-difference-elasticity model~\cite{Seifert1997}.
Below, we will determine an equation for the membrane shape that minimizes the free energy functional associated with Eq.~\eqref{eq:free_energy}.

In-plane motion arising from tangential stresses always keeps the membrane shape fixed, while out-of-plane motion due to normal stresses changes the membrane shape.
Therefore, to find the equilibrium shape of the membrane, one has to first determine the normal stresses acting on its surface.	
This is achieved by considering a virtual displacement of all surface points by an infinitesimal distance $\varphi$ orthogonal to the basis vectors that span the membrane; this is called a \emph{variation}.
Such a variation affects the free energy functional, $F\rightarrow F+\delta F$, by changing membrane curvature, $H + \delta H$, surface area, $S + \delta S$, and surface protein density $m + \delta m$.
Then, one can determine the normal stress, $\sigma_z = \delta F / \delta \varphi$, by a straight-forward calculation involving variational calculus~\cite{Goychuk2019}:
\begin{multline}
	\sigma_z = 
	\tfrac12\kappa\big(H-H_0\big)(H^2 - 4 K + H H_0) 
	+\Delta_S \big[\kappa\, (H-H_0)\big] \\
	+\epsilon \, m\, \big(H - H_\text{opt} \big) \big(H^2 -2K\big)
	+\Delta_S \big[\epsilon\, m \, (H - H_\text{opt}) \big] \\
	+ H \left[m \, \partial_m f_{\text{PP}}(m) - f_{\text{PP}}(m)\right] \, ,
\label{eq:normal_stress}
\end{multline}
where $K$ denotes the Gaussian curvature and $\Delta_S$ the surface Laplacian operator.
The above equation has three main contributions, which we list below sorted by lines:
\begin{enumerate}
	\item stress from bending the membrane away from its intrinsic curvature~\cite{Guckenberger2017,Deserno2015,Zhong1989},
	\item stress from mechanochemical coupling between proteins and membrane, and
	\item stress from protein-protein interactions.
\end{enumerate}
Note that the last line suggests that interactions between proteins can lead to membrane deformations, for example by protein crowding~\cite{Stachowiak2012}.
In mechanical equilibrium, normal stresses on the membrane must vanish, $\sigma_z = 0$.
This yields the shape equation that minimizes the free energy functional associated with Eq.~\eqref{eq:free_energy}.

\subsection{Shape equation for small deformations}

Here, we use the results from Sec.~\ref{sec:arbitrary_deformations} to derive a shape equation that is valid for sufficiently small curvatures.
We assume that intrinsic membrane curvature, $H_0$, and membrane curvature, $H$, are both sufficiently small, and therefore neglect the corresponding nonlinear terms in Eq.~\eqref{eq:normal_stress}.
Furthermore, we assume that direct interactions between proteins do not significantly contribute to membrane deformations, and therefore neglect the third line of Eq.~\eqref{eq:normal_stress}.
Then, the shape equation dramatically reduces to a Laplace equation:
\begin{equation}
	\Delta_S \big[\kappa\, (H-H_0) + \epsilon \, m \, (H - H_\text{opt})\big] = 0 \, ,
\label{eq:shape_equation}
\end{equation}
which can be easily solved by integrating twice.
To that end, we consider a large membrane with fixed curvature, $H|_{\partial S} = H_0$, and vanishing surface protein density, $m|_{\partial S} = 0$, at the boundaries of integration.
The only solution to Eq.~\eqref{eq:shape_equation} that always satisfies these boundary conditions is given by
\begin{equation}
	H = H_0 + (H_\text{opt} - H_0) \frac{\epsilon\, m}{\kappa+\epsilon\, m} \, ,
\end{equation}
which is identical to the expression derived in the main text.
Therefore, we conclude that, as long as the curvature induced by the proteins remains sufficiently small, the results obtained in the main text can be applied to a general spatially extended setting.

\subsection{Tangential forces on the proteins}

In the main text, we have neglected the effects of gradients in membrane conformation and protein density on the distribution of proteins.
When one considers such gradients, one finds that proteins can self-organize on the membrane~\cite{Haselwandter2013, Haselwandter2014, Vahid2016, vanderWel2016, AgudoCanalejo2017, Vahid2017, Vahid2018, Idema2019}.
Specifically, for proteins that \emph{locally force} the membrane to a given shape, one finds that proteins with a symmetric curvature profile repel each other on the membrane, while crescent-shaped proteins can also attract each other~\cite{Schweitzer2015}.
Here, we consider proteins that do not strictly enforce a density-independent local curvature, but \emph{gradually} deform the membrane with increasing protein density.
In that case, one finds attractive interactions between proteins that lead to the accumulation of proteins in regions of preferred curvature, which we will briefly outline in this section.

Tangential displacements of surface points keep the shape of the membrane (principal curvatures and surface area) fixed, while translating the protein density along the surface.
Analogously to out-of-plane motion, see Sec.~\ref{sec:arbitrary_deformations}, in-plane translations can also have an effect on the free energy, which then resulting in effective tangential forces.
We proceed by determining the chemical potential, $\mu_m(m)=\delta F/\delta m$, which is encodes how a protein density variation affects the free energy functional:
\begin{multline}
	\mu_\text{m}(m) = 
	\tfrac{1}{2} \, \epsilon \, 
	(H - H_\text{opt})^2 \\
	+ \epsilon \, (H - H_\text{opt}) \, m \, \partial_m H 
	+ \partial_m f_{\text{PP}}(m) \, .
\label{eq:chempot_surface}
\end{multline}
Here, the membrane curvature depends on the protein density, as the mechanical degrees of freedom are assumed to relax instantaneously.
Each protein that moves in the chemical potential, Eq.~\eqref{eq:chempot_surface}, experiences an in-plane force given by $\mathbf{f} = -\boldsymbol\nabla_S \mu_\text{m} (m)$.
This becomes relevant when one considers protein self-organization on the surface  (agglomeration towards regions of preferred curvature) in addition to protein recruitment, but has no effect on the binding kinetics itself.